\documentclass[journal]{IEEEtran}

\usepackage{ifpdf}
\usepackage{url}
\ifpdf
\else
\fi

\usepackage{cite}
\usepackage{balance}
\usepackage{bm,comment,color}

\ifCLASSINFOpdf
  \usepackage[pdftex]{graphicx}
  \graphicspath{{../pdf/}{../jpeg/}}
  \DeclareGraphicsExtensions{.pdf,.jpeg,.png}
\else
  \usepackage[dvips]{graphicx}
  \graphicspath{{../eps/}}
  \DeclareGraphicsExtensions{.eps}
\fi
\usepackage{amsmath}
\usepackage{amsthm}
\newtheorem*{remark}{Remark}

\usepackage{algorithm}
\usepackage{algpseudocode}
\algtext*{EndWhile}
\algtext*{EndIf}
\algtext*{EndFor}

\usepackage{array}
\usepackage{amsfonts} 
\usepackage{amssymb}
\usepackage{esint} 
\usepackage{units}
\usepackage{multirow}
\usepackage{comment}

\usepackage{color}
\newcommand{\red}[1]{{\color{red}{#1}}} 
\newcommand{\cyan}[1]{{\color{cyan}{#1}}}

\newcommand{\orange}[1]{{\color{orange}{#1}}} 
\newcommand{\violet}[1]{{\color{violet}{#1}}} 

\usepackage{pgfplots}
\usepackage{tikz}
\usetikzlibrary{calc}
\makeatletter
\newcommand{\gettikzxy}[3]{%
  \tikz@scan@one@point\pgfutil@firstofone#1\relax
  \edef#2{\the\pgf@x}%
  \edef#3{\the\pgf@y}%
}
\usetikzlibrary{spy,backgrounds}
\pgfplotsset{compat=newest}
\usetikzlibrary{plotmarks}
\usetikzlibrary{arrows.meta}
\usepgfplotslibrary{patchplots}
\usepackage{grffile}
\newlength\fheight 
\newlength\fwidth 
\usepgfplotslibrary{fillbetween}

\usepackage{acronym}
\usepackage{tabu,longtable}

\ifCLASSOPTIONcompsoc
 \usepackage[caption=false,font=normalsize,labelfont=sf,textfont=sf]{subfig}
\else
 \usepackage[caption=false,font=footnotesize]{subfig}
\fi
\usepackage{stfloats}
\usepackage[hidelinks]{hyperref}
\usepackage{xcolor}
\long\def\comment#1{}

\DeclareMathOperator*{\argmin}{arg\,min}

\newfont{\bbb}{msbm10 scaled 700}

\newcommand{\hthickline}{\noalign{\hrule height 0.80pt}}

\newfont{\bb}{msbm10 scaled 1100}

\newcommand{\hermit}{\mathsf{H}}

\newcommand{\av}{{\bf a}}
\newcommand{\bv}{{\bf b}}

\newcommand{\dv}{{\bf d}}
\newcommand{\ev}{{\bf e}}

\newcommand{\hv}{{\bf h}}

\newcommand{\lv}{{\bf l}}

\newcommand{\ov}{{\bf o}}
\newcommand{\pv}{{\bf p}}

\newcommand{\sv}{{\bf s}}
\newcommand{\tv}{{\bf t}}

\newcommand{\xv}{{\bf x}}

\newcommand{\zv}{{\bf z}}

\newcommand{\avr}{\av_\text{R}}

\newcommand{\Am}{{\bf A}}
\newcommand{\Bm}{{\bf B}}

\newcommand{\Dm}{{\bf D}}

\newcommand{\Hm}{{\bf H}}

\newcommand{\Jm}{{\bf J}}

\newcommand{\Nm}{{\bf N}}
\newcommand{\Om}{{\bf O}}

\newcommand{\Qm}{{\bf Q}}
\newcommand{\Rm}{{\bf R}}
\newcommand{\Sm}{{\bf S}}
\newcommand{\Tm}{{\bf T}}

\newcommand{\Xm}{{\bf X}}
\newcommand{\Ym}{{\bf Y}}
\newcommand{\Zm}{{\bf Z}}

\newcommand{\deltav}{\hbox{\boldmath$\delta$}}
\newcommand{\etav}{\hbox{\boldmath$\eta$}}

\newcommand{\muv}{\hbox{\boldmath$\mu$}}

\newcommand{\phiv}{\hbox{\boldmath$\phi$}}
\newcommand{\psiv}{\hbox{\boldmath$\psi$}}

\newcommand{\omegav}{\hbox{\boldmath$\omega$}}

\newcommand{\boldone}{{ {\boldsymbol{1}} }}

\newcommand{\varphiv}{\hbox{\boldmath$\varphi$}}
\newcommand{\vpv}{\boldsymbol{\varphi}}

\newcommand{\Sigmam}{\hbox{\boldmath$\Sigma$}}

\newcommand{\Omegam}{\hbox{\boldmath$\Omega$}}
\newcommand{\Xim}{\hbox{\boldmath$\Xi$}}

\newcommand{\trace}{{\hbox{tr}}}

\renewcommand{\arg}{{\hbox{arg}}}

\newcommand{\herm}{{\sf H}}

\acrodef{3gpp}[3GPP]{3rd Gneration Partnership Project}
\acrodef{ad}[AD]{autonomous drive}
\acrodef{adas}[ADAS]{advanced driver assistance system}
\acrodef{aoa}[AOA]{angles-of-arrival}
\acrodef{aod}[AOD]{angles-of-departure}
\acrodef{aosa}[AOSA]{array-of-subarray}
\acrodef{bs}[BS]{base station}
\acrodef{bse}[BSE]{beam squint effect}
\acrodef{cdf}[CDF]{cumulative distribution function}
\acrodef{coa}[COA]{curvature of arrival}
\acrodef{crb}[CRB]{Cram\'er-Rao bound}
\acrodef{dbscan}[DBSCAN]{density-based spatial clustering of applications with noise}
\acrodef{dft}[DFT]{discrete Fourier transform}
\acrodef{dl}[DL]{deep learning}
\acrodef{elaa}[ELAA]{extremely-large antenna array}
\acrodef{ff}[FF]{far field}
\acrodef{fim}[FIM]{Fisher information matrix}
\acrodef{gnss}[GNSS]{global navigation satellite system}
\acrodef{gps}[GPS]{global positioning system}
\acrodef{imu}[IMU]{inertial measurement unit}
\acrodef{ip}[IP]{incidence point}
\acrodef{its}[ITS]{intelligent transport system}
\acrodef{kld}[KLD]{Kullback–Leibler divergence}
\acrodef{las}[L\&S]{localization and sensing}
\acrodef{los}[LOS]{line-of-sight}
\acrodef{mae}[MAE]{mean absolute value}
\acrodef{map}[MAP]{maximum a posteriori}
\acrodef{mimo}[MIMO]{multiple-input-multiple-output}
\acrodef{siso}[SISO]{single-input-single-output}
\acrodef{simo}[SIMO]{single-input-multiple-output}
\acrodef{miso}[MISO]{multiple-input-single-output}
\acrodef{ml}[ML]{machine learning}
\acrodef{mle}[MLE]{maximum likelihood estimator}
\acrodef{mm}[MM]{mismatched model}
\acrodef{mmwave}[mmWave]{millimeter wave}
\acrodef{mpc}[MPC]{multipath component}
\acrodef{nlos}[NLOS]{non-line-of-sight}
\acrodef{nf}[NF]{near field}
\acrodef{nr}[NR]{new radio}
\acrodef{ofdm}[OFDM]{orthogonal frequency division multiplexing}
\acrodef{pbd}[PBD]{partial blockage detection}
\acrodef{prs}[PRS]{positioning reference signal}
\acrodef{psd}[PSD]{power spectral density}
\acrodef{pss}[PSS]{primary synchronization signal}
\acrodef{rcs}[RCS]{radar cross section}

\acrodef{rx}[RX]{receiver}
\acrodef{rf}[RF]{radio frequency}
\acrodef{rfc}[RFC]{radio frequency chain}

\acrodef{ris}[RIS]{reconfigurable intelligent surface}
\acrodef{rss}[RSS]{received signal strength}
\acrodef{rtk}[RTK]{real-time kinematic}
\acrodef{rtt}[RTT]{round-trip-time}
\acrodef{sa}[SA]{sub-array}
\acrodef{simo}[SIMO]{single-input-multiple-output}
\acrodef{slam}[SLAM]{simultaneous localization and mapping}
\acrodef{snr}[SNR]{signal-to-noise-ratio}
\acrodef{sns}[SNS]{spatial non-stationarity}
\acrodef{sp}[SP]{scattering point}
\acrodef{ssb}[SSB]{synchronization signal/physical broadcast channel block}
\acrodef{swm}[SWM]{spherical wave model}
\acrodef{tx}[TX]{transmitter}
\acrodef{tdd}[TDD]{time division duplex}
\acrodef{tdoa}[TDOA]{time-difference-of-arrival}
\acrodef{thz}[THz]{Terahertz}
\acrodef{tm}[TM]{true model}
\acrodef{toa}[TOA]{time-of-arrival}
\acrodef{ue}[UE]{user equipment}
\acrodef{ura}[URA]{uniform rectangular array}
\acrodef{va}[VA]{virtual anchor}

\makeatletter
\def\ps@IEEEtitlepagestyle{%
\def\@oddfoot{\mycopyrightnotice}%
\def\@evenfoot{}%
}
\def\mycopyrightnotice{%
{
} 
\gdef\mycopyrightnotice{}
}

\makeatletter
\let\old@ps@headings\ps@headings
\let\old@ps@IEEEtitlepagestyle\ps@IEEEtitlepagestyle
\def\confheader#1{
\def\@oddhead{\strut\hfill#1\hfill\strut}%
}
\makeatother
\confheader{\scriptsize \shortstack{This work has been submitted to the IEEE for possible publication.  Copyright may be transferred without notice, after which this version may no longer be accessible.}}

\begin{document} 

\bstctlcite{IEEEexample:BSTcontrol}

\title{Multi-RIS-Enabled 3D Sidelink Positioning}

\author{
Hui~Chen,~\IEEEmembership{Member,~IEEE},
Pinjun~Zheng,
Musa~Furkan~Keskin,~\IEEEmembership{Member,~IEEE},
Tareq~Al-Naffouri,~\IEEEmembership{Senior~Member,~IEEE}
and~Henk~Wymeersch,~\IEEEmembership{Senior~Member,~IEEE}

\thanks{H.~Chen, M. F. Keskin and H.~Wymeersch are with the Department of Electrical Engineering, Chalmers University of Technology, 412 58 Gothenburg, Sweden (Email: {hui.chen; furkan; henkw}@chalmers.se). P.~Zheng and  T.~Y.~Al-Naffouri are with the Division of Computer, Electrical and Mathematical Science \& Engineering, King Abdullah University of Science and Technology (KAUST), Thuwal, 23955-6900, KSA. (Email: \{pinjun.zheng; tareq.alnaffouri\}@kaust.edu.sa). H.~Chen and P. Zheng are co-first authors; they contributed equally to this paper.}



\thanks{This work was supported, in part, by the European Commission through the EU H2020 RISE-6G project under grant 101017011, by the King Abdullah University of Science and Technology (KAUST) Office of Sponsored Research (OSR) under Award No. ORACRG2021-4695, and by the 6G-Cities project from Chalmers.}

}


\maketitle


\begin{abstract}
Positioning is expected to support communication and location-based services in the fifth/sixth generation (5G/6G) networks. With the advent of reflective reconfigurable intelligent surfaces (RISs), radio propagation channels can be controlled, making high-accuracy positioning and extended service coverage possible.
However, the passive nature of the RIS requires a signal source such as a base station (BS), which limits the positioning service in extreme situations, such as tunnels, dense urban areas, or complicated indoor scenarios where 5G/6G BSs are not accessible. In this work, we show that with the assistance of (at least) two RISs and sidelink communication between two user equipments (UEs), the absolute positions of these UEs can be estimated in the absence of BSs. A two-stage 3D sidelink positioning algorithm is proposed, benchmarked by the derived Cram\'er-Rao bounds. The effects of multipath and RIS profile designs on positioning performance are evaluated, and localization analyses are performed for various scenarios. Simulation results demonstrate the promising positioning accuracy of the proposed BS-free sidelink communication system. Additionally, we propose and evaluate several solutions to eliminate potential blind areas where positioning performance is poor, such as removing clock offset via round-trip communication, adding geometrical prior or constraints, as well as introducing more RISs.
\end{abstract}


\IEEEpeerreviewmaketitle
\acresetall 

\section{Introduction}
\label{sec:intro}
{Knowing the position of a mobile device is crucial in various scenarios, starting from emergency services (e.g., disaster rescue) and then extended to daily location-based applications (e.g., navigation and augmented reality)~\cite{del2017survey}. 
\Ac{gnss} is one of the most successful positioning systems, which is highly effective in outdoor scenarios but has persistent challenges in dense urban areas and indoor environments.
With the increased frequency and bandwidth of the \ac{mmwave}/\ac{thz} systems in the fifth/sixth generation (5G/6G) systems, promising positioning solutions are emerging to complement \ac{gnss}, and also making communication and positioning integrated and beneficial to each other~\cite{de2021convergent,chen2022tutorial}.}
Position information can be extracted from channel estimation using radio signals, which can further assist communication with handover~\cite{santi2022location}, and re-establishment of communication links~\cite{chen2022tutorial}. Such integration provides versatile 5G/6G radio systems that can support both communications and positioning functions without introducing extra infrastructure deployments. 


Positioning in the 5G new radio (NR) has been studied in TR38.855~\cite{tr38855} and more recently in TR38.895~\cite{tr38895}, with initial efforts being carried out in both academia and industry. Based on existing \ac{mmwave} positioning research works, huge potential has been shown in angle-based positioning~\cite{shahmansoori2017position}, multipath resolvability~\cite{li2019massive}, positioning under mobility~\cite{keykhosravi2022ris}, and 6D positioning scenarios~\cite{nazari2022mmwave}. Verification and evaluation of the onsite positioning systems have also been carried out with 5G base stations in  indoor~\cite{yammine2021experimental, gao2022toward} and outdoor scenarios~\cite{ge2022experimental}. However, no existing works have reported 5G positioning with comparable performance reported in theoretical analysis or expected in future use cases. The model mismatch (e.g., caused by hardware impairment~\cite{chen2023modeling}, the effect of multipath~\cite{dardari2021nlos}, erroneous motion model~\cite{guerra2021near}) and harsh propagation channels~\cite{behravan2022positioning} constitute major factors that prevent the radio positioning system from achieving high-accuracy performance. 
These factors cause errors in channel parameter estimation and further affect the positioning performance, especially for UEs located far away from the BSs, where \ac{snr} is low and angle estimation error propagates with distance.
Laying out a denser network with more active anchors can mitigate the above-mentioned positioning errors. However, network deployment cost increases with densification (especially for positioning where usually multiple \acp{bs} are needed at the same time), requiring new enablers to accomplish positioning tasks.

One of the most promising enablers is sidelink communication (or device-to-device communication),  introduced in 3GPP Release 12~\cite{lin2014overview}, and more recently standardized in Release 16~\cite{tr38885} to support FR1 and the \ac{mmwave} range FR2. With direct communication between devices, cooperative positioning is possible, which reduces the requirement for densely deployed BSs. In general, relative position information between each device/vehicle can be obtained in a cooperative positioning network given a sufficient number of vehicles~\cite{chukhno2021d2d}. With an anchor provided in a global coordinate system, the true positions of all the devices can be obtained. Moreover, the sidelink can also be implemented in partial coverage and out-of-coverage areas for positioning, where the relative location will be beneficial to vehicles in various applications such as platooning, collision avoidance, and so on~\cite{garcia2021tutorial}.

Another promising technology that has been studied extensively for positioning (yet not standardized) is \ac{ris}~\cite{liu2021reconfigurable, wymeersch2020radio,bjornson2022reconfigurable}.\footnote{Throughout this work, we consider passive RISs due to their advantage of low power consumption and low deployment cost. However, other types of RIS also exist, such as active RIS~\cite{zhang2022active}, hybrid RIS~\cite{schroeder2022two}, and simultaneous transmissive and receiving (STAR) RIS~\cite{xu2021star}, which are left for future work.} RISs consist of configurable elements with the ability to reshape the channel by changing the phase of the incident signals. For communication, RISs are able to provide improved \ac{snr}, reduced interference, and extended coverage under blockage. From the positioning point of view, RISs can work as additional passive anchors and provide high-resolution angular information by virtue of a large number of RIS elements. With the assistance of RIS, various positioning scenarios are created, from the simplest scenario where a UE can be localized in a \ac{siso} system~\cite{keykhosravi2022ris} to joint localization and mapping with multiple \acp{ris}~\cite{lin2021channel}. In bi-static and multi-static sensing scenarios, when an object is equipped with a RIS, the object can be passively localized with transmitter and receiver anchors~\cite{ghazalian2022bi}. More recent works show joint RIS calibration and UE positioning can be performed simultaneously within a \ac{simo} system, providing a practical solution for RIS calibration~\cite{lu2022joint, zheng2023jrcup}. All of these works have shown a huge potential for RIS in 5G/6G positioning.

Both sidelink communication and RIS are promising enablers for 5G/6G systems, which have been separately studied in most of the works. The discussion on combining these two technologies for low-latency and high-reliability communication has been discussed in~\cite{gu2022intelligent}, while positioning works appear quite recently~\cite{chen2023riss,balasubramanian2022reconfigurable} with {a major feature of no \acp{bs} being involved, resulting in the \ac{tx} and \ac{rx} both with unknown positions.} In~\cite{chen2023riss}, sidelink positioning with RISs is discussed at a high level {with localization and sensing scenarios, architectures, and protocols being discussed}. {Regarding the technical contributions}, the work in~\cite{balasubramanian2022reconfigurable} requires the cooperation of multiple UEs {with only time-of-arrival information being considered, and self-localization has been studied in~\cite{kim2022ris} where the \ac{ue} is equipped with a full-duplex array. However, the former case did not benefit from the high angular resolution of RISs, and the latter scenario requires extra hardware cost to upgrade half-duplex arrays into (radar-like) full-duplex arrays. More recent work shows that cooperative localization can benefit from spatial frequency estimation with a single RIS. However, at least three UEs are required, and the coordination of multi-UE transmission increases positioning overhead~\cite{ammous2023zero}.}

To the best of our knowledge, this is the first technical work that discusses the multi-RIS-enabled 3D sidelink positioning problem, {which aims to locate both the transmitter and the receiver using one-way pilot signal transmission} without full-duplex hardware requirements. We will show that \textbf{{with a sufficient number of \acp{ris} (at least two) involved, the 3D absolute positions of two single-antenna UEs can be estimated using sidelink communication in the absence of BSs}}, making ubiquitous positioning possible.
{Such an application can be used in partial coverage and out-of-coverage scenarios (such as tunnels or rural areas) without any BSs involved~\cite{tr23700, tr38845, harounabadi2021v2x}.
By relying on the deployment of quasi-passive RISs and the cooperation of UEs/vehicles via sidelink communication, localization tasks can be accomplished with low-cost systems.}

In this work, we consider a 3D \ac{siso} sidelink communication scenario with two UEs and several passive RIS anchors. The contributions of this work can be summarized as follows:
\begin{itemize}
    \item We formulate the problem of multi-RIS enabled 3D \ac{siso} sidelink positioning. In this scenario, the RISs (at least two) work as passive anchors with known positions and orientations. With sidelink communication, the 3D positions (absolute positions with respect to the global coordinate system) of both UEs and the clock offset between them can be estimated. {Note that this problem is fundamentally different from previous positioning works as we do not assume the state of one of the \ac{tx} or \ac{rx} to be known as in~\cite{keykhosravi2022ris}, and we do not assume a full-duplex transceiver is implemented as in~\cite{kim2022ris}.}
    \item We derive the \acp{crb} for both channel parameter estimation and positioning, which serve several purposes: a) to benchmark the proposed positioning algorithms; b) to evaluate the different designs of RIS profiles; c) to provide guidelines on blind areas (where positioning task cannot be completed) evaluation and anchor deployment optimization.
    \item We adopt a time-orthogonal RIS profile design scheme to assist channel estimation by differentiating the LOS path, and each of the RIS paths from each other. With this scheme, we design positioning-oriented RIS profiles based on directional and derivative codebooks from prior UE information, which can be further improved with power control.
    \item We develop a low-complexity channel parameter estimation to obtain the delays and spatial frequencies (separate estimation of \ac{aoa} and \ac{aod} is not possible in this scenario due to inherent ambiguity, which will be described in Section~\ref{sec:problem_statement}). Based on the delay and spatial frequency estimates from multiple RISs, a 3D-search positioning algorithm is developed to estimate the 3D positions of both UEs and the clock offset between them. In addition, maximum likelihood estimators for channel parameter estimation and positioning are also formulated for refining results.
    \item Extensive simulations are carried out to show the effectiveness of the derived performance analysis and the proposed algorithm. The effects of multipath and RIS profile designs on positioning performance are evaluated. Several RIS deployment strategies (e.g., placed on one side or both sides of the road), and further sidelink positioning system designs are suggested.
\end{itemize}

The structure of this paper is organized as follows. Section~\ref{sec_system_model} discusses the system model, based on which problem formulation will be described. 
The performance analysis, including the lower bounds for channel parameters and position estimation, is provided in Section~\ref{sec_performance_analysis}. Section~\ref{sec_methodology} details the methodology of the RIS profile design and positioning algorithm. Simulation results are presented in Section~\ref{sec_simulation_results}, followed by the conclusion of this work in Section~\ref{sec_conclusion}.

\textit{Notations and Symbols:} Italic letters denote scalars (e.g. $a$), bold lower-case letters denote vectors (e.g. $\av$), and bold upper-case letters denote matrices (e.g. $\Am$). $(\cdot)^\top$, $(\cdot)^\herm$, $(\cdot)^*$, $(\cdot)^{-1}$, $\trace(\cdot)$, and $\Vert{\cdot}\Vert$ represent the transpose, Hermitian transpose, complex conjugate, inverse, trace, and $\ell$-2 norm operations, respectively; $\Am \odot \Bm$, $\Am \otimes \Bm$, $\av \circ \bv$ are the Hadamard product, Kronecker product, and outer product, respectively; $[\cdot,\ \cdot,\ \cdots, \cdot]^\top$ denotes a column vector; $[\cdot]_{i,j}$ is the element in the $i$-th row, $j$-th column of a matrix, and $[\cdot]_{a:b,c:d}$ is the submatrix constructed from the $a$-th to the $b$-th row, and the $c$-th to $d$-th column of a matrix; $\mathrm{Re}\{a\}$ extracts the real part of a complex variable, $\angle(a)$ returns the phase of a complex number $a$; $\bm{1}_{N}$ denotes an $N\times 1$ all ones vector, and $\mathbf{I}_{N}$ denotes a size-$N$ identity matrix.


\begin{figure*}[t]
    \centering
    \begin{tikzpicture}
    \node (image) [anchor=south west]{\includegraphics[width=.65\linewidth]{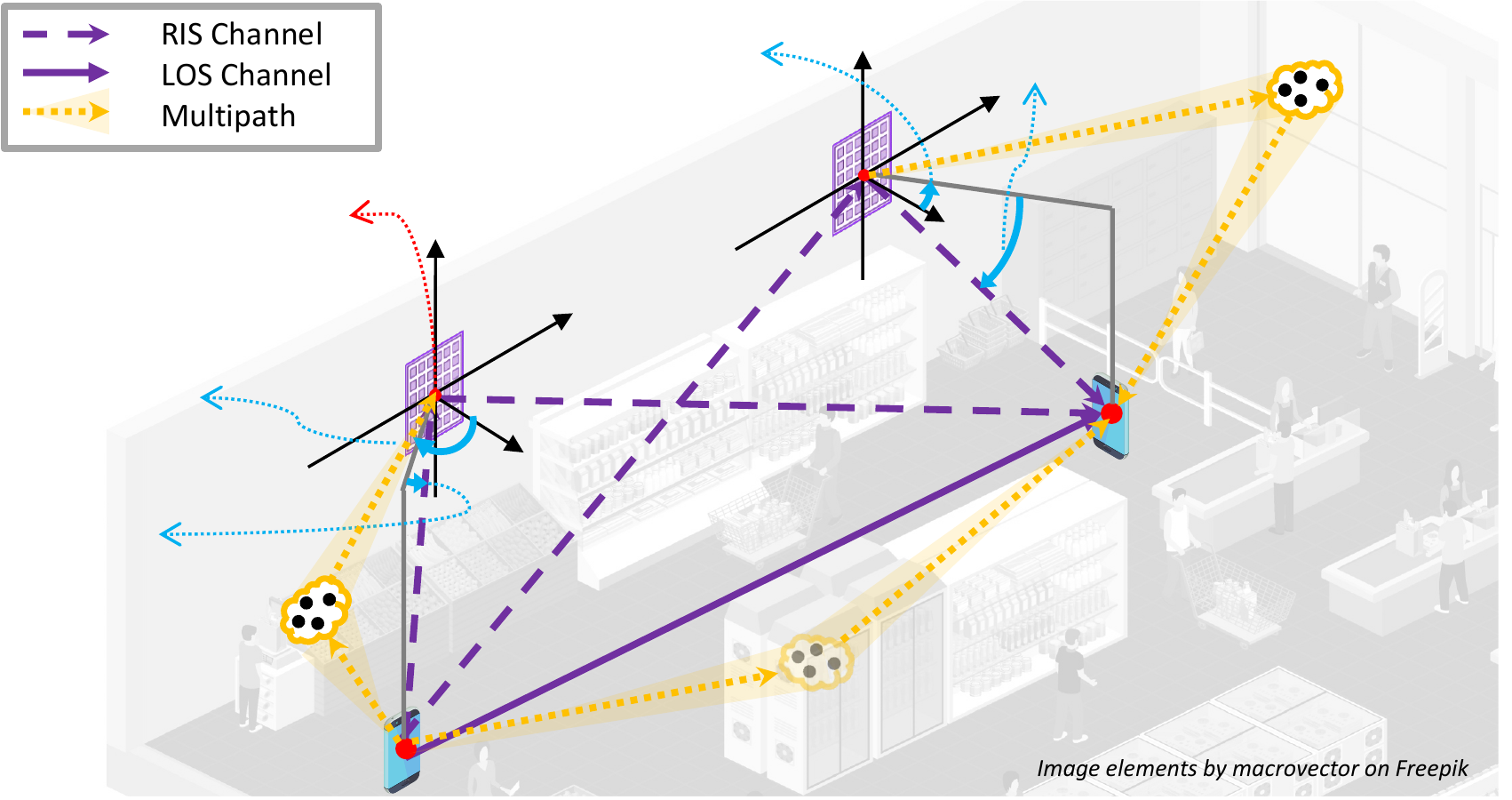}};
    \gettikzxy{(image.north east)}{\ix}{\iy};
    \node at (0.37*\ix,0.5*\iy)[rotate=0,anchor=north]{\small{$x$}};
    \node at (0.405*\ix,0.685*\iy)[rotate=0,anchor=north]{\small{$y$}};
    \node at (0.295*\ix,0.77*\iy)[rotate=0,anchor=north]{\small{$z$}};
    \node at (0.15*\ix,0.8*\iy)[rotate=0,anchor=north]{\tiny{\red{\shortstack{Origin of the Global \\Coordinate System: \\ $[0,0,0,]^\top \unit[]{m}$}}}};
    \node at (0.25*\ix,0.62*\iy)[rotate=0,anchor=north]{\small{\red{$\pv_1$}}};
    \node at (0.52*\ix,0.88*\iy)[rotate=0,anchor=north]{\small{\red{$\pv_2$}}};
    \node at (0.21*\ix,0.13*\iy)[rotate=0,anchor=north]{\small{\red{$\pv_\text{T}$ ?}}};
    \node at (0.80*\ix,0.55*\iy)[rotate=0,anchor=north]{\small{\red{$\pv_\text{R}$ ?}}};
    \node at (0.09*\ix,0.4*\iy)[rotate=0,anchor=north]{\small{\cyan{$\theta_{\text{A},1}$}}};
    \node at (0.1*\ix,0.58*\iy)[rotate=0,anchor=north]{\small{\cyan{$\phi_{\text{A},1}$}}};
    \node at (0.68*\ix,0.97*\iy)[rotate=0,anchor=north]{\small{\cyan{$\theta_{\text{D},2}$}}};
    \node at (0.47*\ix,0.98*\iy)[rotate=0,anchor=north]{\small{\cyan{$\phi_{\text{D},2}$}}};
    \node at (0.54*\ix,0.42*\iy)[rotate=0,anchor=north]{\small{\violet{$\Hm_\text{U}$}}};
    \node at (0.34*\ix,0.47*\iy)[rotate=0,anchor=north]{\small{\violet{$\Hm_\text{R,1}$}}};
    \node at (0.57*\ix,0.68*\iy)[rotate=0,anchor=north]{\small{\violet{$\Hm_\text{R,2}$}}};
    \node at (0.13*\ix,0.28*\iy)[rotate=0,anchor=north]{\small{\orange{$\Hm_\text{R,MP,1}$}}};
    \node at (0.87*\ix,0.72*\iy)[rotate=0,anchor=north]{\small{\orange{$\Hm_\text{R,MP,2}$}}};
    \node at (0.57*\ix,0.17*\iy)[rotate=0,anchor=north]{\small{\orange{$\Hm_\text{U,MP}$}}};
    \end{tikzpicture}
    \caption{Illustration of multi-RIS-enabled 3D sidelink positioning. With the help of multiple (at least two) RIS anchors, the positions of both UEs (with respect to the global coordinate system) and the clock offset between them can be estimated through a one-way sidelink communication.}
    \label{fig_illustration}\vspace{-5mm}
\end{figure*}
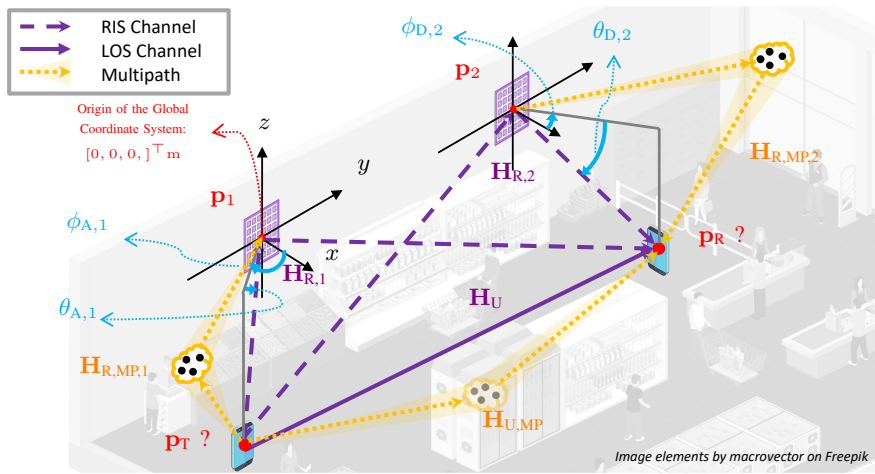

\section{System Model}
\label{sec_system_model}
In this section, we describe the geometry model, signal model, and problem statement of the considered multi-RIS-enabled 3D sidelink positioning.
\subsection{Geometry Model}
We consider a 3D \ac{siso} scenario with $L>1$ RISs and two unsynchronized single-antenna \acp{ue}, where the 3D positions of both UEs need to be estimated via sidelink communication, as shown in Fig.~\ref{fig_illustration}. The \ac{tx} and the \ac{rx} \acp{ue} are located at $\pv_{\text{T}}$, $\pv_{\text{R}}$ $\in \mathbb{R}^3$, respectively. The positions (array centers) and orientations of $L$ \acp{ris} are denoted by $\pv_1, \ldots, \pv_L$ $\in \mathbb{R}^{3}$, 
and Euler angle vectors $\ov_1, \ldots, \ov_L \in \mathbb{R}^{3}$ {(which can be mapped into rotation matrices $\Rm_1, \ldots, \Rm_L$ that belong to the {special orthogonal group in three dimensions} denoted $\text{SO(3)}$~\cite{chen2022tutorial})}, respectively. 
For simplicity, we assume all the \acp{ris} consist of $N = N_1\times N_2$ RIS elements with $N_1$ and $N_2$ as the number of rows and columns, respectively. In addition, without loss of generality, all the RIS elements are located on the Y-Z plane of each RIS's local coordinate system with the $n$-th element located at $\zv_{n} = [0, z_{n,2}, z_{n,3}]$. The \ac{aoa} $\vpv_{\text{A},\ell}$ at the $\ell$-th RIS from the \ac{tx} UE and the \ac{aod} $\vpv_{\text{D},\ell}$ from the same RIS to the \ac{rx} UE  can then be expressed as
\begin{align}
    \vpv_{\text{A},\ell} & = 
    \begin{bmatrix}
        \phi_{\text{A},\ell}\\
        \theta_{\text{A},\ell}
    \end{bmatrix}=
    \begin{bmatrix}
        \arctan2( t_{\text{T},\ell,2}, t_{\text{T},\ell,1})\\
        \arcsin( t_{\text{T},\ell,3})
    \end{bmatrix},
    \label{eq:aoa}
    \\
    \vpv_{\text{D},\ell} & = 
    \begin{bmatrix}
        \phi_{\text{D},\ell}\\
        \theta_{\text{D},\ell}
    \end{bmatrix}=
    \begin{bmatrix}
        \arctan2( t_{\text{R},\ell,2}, t_{\text{R},\ell,1})\\
        \arcsin( t_{\text{R},\ell,3})
    \end{bmatrix},
    \label{eq:aod}
\end{align}
where $\phi$ and $\theta$ are the azimuth and elevation angles, respectively. Let $\tv_{\text{T},\ell} = [ t_{\text{T},\ell,1},  t_{\text{T},\ell,2},  t_{\text{T},\ell,3}]^\top$ and $\tv_{\text{R},\ell} = [t_{\text{R},\ell,1},  t_{\text{R},\ell,2},  t_{\text{R},\ell,3}]^\top$ denote the direction vectors in the local coordinate system of the $\ell$-th RIS to the TX and RX, respectively.  These vectors can be expressed using global positions $\pv_\text{T}$, $\pv_\text{R}$, $\pv_\ell$ and rotation matrix $\Rm_\ell$ as

\begin{align}
    \tv_{\text{T},\ell} & =  \Rm_\ell^{-1}\frac{\pv_\text{T} - \pv_\ell}{\Vert \pv_\text{T} - \pv_\ell \Vert}=
    \begin{bmatrix}        \cos(\phi_{\text{A},\ell})\cos(\theta_{\text{A},\ell})
    \\
    \sin(\phi_{\text{A},\ell})\cos(\theta_{\text{A},\ell})\\
    \sin(\theta_{\text{A},\ell})\\
    \end{bmatrix}
    ,
    \label{eq:tv_T}
    \\
    \tv_{\text{R},\ell} & =  \Rm_\ell^{-1}\frac{\pv_\text{R} - \pv_\ell}{\Vert \pv_\text{R} - \pv_\ell \Vert}=
    \begin{bmatrix}        \cos(\phi_{\text{D},\ell})\cos(\theta_{\text{D},\ell})
    \\
    \sin(\phi_{\text{D},\ell})\cos(\theta_{\text{D},\ell})\\
    \sin(\theta_{\text{D},\ell})\\
    \end{bmatrix}.
    \label{eq:tv_R}
\end{align}
{We also assume the existence of $M=\sum_{\ell=0}^L M_\ell$ \acp{mpc} with $M_0$ TX-{\ac{sp}}-RX \acp{mpc}, and $M_\ell$ \acp{mpc} in the $\ell$-th \ac{ris} channel (in the forms of TX-SP-RIS-RX or TX-RIS-SP-RX). The corresponding \ac{aod}, \ac{aoa}, and direction vectors at the \ac{ris} with respect to the \ac{sp} $\pv_{\ell, m}$ can be defined similarly based on \eqref{eq:aoa} to \eqref{eq:tv_R}.}

\subsection{Signal Model}
{Assume $K$ subcarriers are adopted in a wideband system during the sidelink communication}, and $G$ \ac{ofdm} symbols are sent during the coherence time. The received signal block $\Ym\in\mathbb{C}^{K\times G}$ can be formulated as
\begin{equation}
        \Ym  = {\Ym_\text{U}} + {\Ym_{\text{R}}} + \Nm, 
\label{eq:signal_model}
\end{equation}
where $\Nm\in \mathbb{C}^{K\times G}$ is the additive white Gaussian noise matrix with each element $n_{k,g} \sim \mathcal{CN}(0, \sigma_n^2)$ and $\sigma_n^2 = WN_0$ depending on the bandwidth $W$ and the noise power spectral density (PSD) $N_0$, $\Ym_\text{U}$ and $\Ym_\text{R}$ are the received signal matrix of the uncontrollable paths and RIS paths. 
{We adopt a geometric channel model (e.g., the Saleh-Valenzuela model~\cite{saleh1987statistical} used for mmWave and THz systems) to account for the multipath effect~\cite{pan2022overview,zhou2022channel}, and the received signals can be modeled as}
\begin{align}
    \Ym_{\text{U}} 
    & = (\Hm_{\text{U}} + \Hm_{\text{U,MP}})\odot \Xm \label{eq:channel_LOS_MP}
    \\
    & = \Big(\underbrace{{\rho_{0}}\Dm(\tau_{0})}_{\text{LOS channel}}+ \underbrace{\sum_{m=1}^{M_0}\rho_{0,m}\Dm(\tau_{0,m})}_{\text{uncontrolled multipath channel}}\Big) \odot \Xm, \notag
    \\
    \Ym_{\text{R}} 
    & = \sum_{\ell=1}^{L}(\Hm_{\text{R},\ell} + \Hm_{\text{R,MP},\ell})\odot \Xm 
    \label{eq:channel_RIS_MP}
    \\
    & = \sum_{\ell=1}^L \Big(\underbrace{\rho_{\ell}\Dm(\tau_\ell)\odot\Am_\ell(\psiv_\ell)}_{\ell\text{-th RIS channel}} \notag
    \\
    & + \underbrace{\sum_{m=1}^{M_{\ell}}{\rho_{\ell,m}}\Dm(\tau_{\ell,m})\odot\Am_\ell(\psiv_{\ell,m})}_{\ell \text{-th RIS multipath channel}}\Big)\odot \Xm .\notag
\end{align}

Here, $\Hm_\text{U}$, $\Hm_\text{U, MP}$, $\Hm_\text{R}$  and $\Hm_\text{R, MP}$ are the channel matrix of the LOS path, uncontrolled multipath RIS path, and RIS multipath, respectively. The pilot signal matrix $\Xm$ is defined as
\begin{equation}
    \Xm = \sqrt{P}\xv \deltav^\top \in \mathbb{C}^{K\times G}, \ \ \deltav = [\delta_1, \ldots, \delta_G]^\top\ \ (\Vert \deltav \Vert= \sqrt{G}),
    \label{eq_transmission_matrix}
\end{equation}
where $\xv \in \mathbb{C}^K$ ($|x_k| = 1$) represents the transmitted symbols for $K$ subcarriers, and the transmission power of the $g$-th transmission is $\delta_g^2 P$, where $\deltav = \bm{1}_G$ indicates a constant transmit power during $G$ transmissions. Here, we use the same $\xv$ for all the $G$ transmissions for simplicity, and the aim of introducing $\deltav$ is to implement a constrained power control for each transmission/RIS beam to enhance positioning accuracy with the same total transmission power, which will be detailed in~Section~\ref{sec_methodology}.
{The complex channel gains of the \ac{los} path, $m$-th uncontrolled multipath, $\ell$-th RIS path and the $m$-th multipath involving the $\ell$-th RIS are denoted as $\rho_{0}$, $\rho_{0, m}$, $\rho_{\ell}$, and $\rho_{\ell, m}$, respectively.} The subscripts also apply to the signal propagation delays of different paths such as $\tau_{0}$, $\tau_{0, m}$, $\tau_{\ell}$, $\tau_{\ell, m}$.

{The delay matrix $\Dm(\tau) = \dv(\tau) \mathbf{1}_G^\top \in \mathbb{C}^{K\times G}$ contains the delay information of a specific signal propagation path across different subcarriers as\footnote{We assume the movement within the coherence time is negligible and hence the delay at the $k$-th subcarrier is identical across different transmissions.}}
\begin{equation}
    [\Dm(\tau)]_{k,g} = d_k(\tau) = e^{-j2\pi k \Delta_f \tau},
    \label{eq:delay_matrix}
\end{equation}
with $\Delta_f=W/K$ as the subcarrier spacing. The delay for the LOS channel $\tau_0$ and for the $\tau_\ell$ RIS channel can be expressed as
\begin{align}
    \tau_{0} & = \frac{d_{0} + B}{c}= \frac{{\Vert \pv_\text{T} - \pv_\text{R} \Vert} + B}{c},
    \\
    \tau_{\ell} & =\frac{d_{\text{T},\ell} + d_{\text{R},\ell} + B}{c}=\frac{{\Vert \pv_\text{T} - \pv_{\ell} \Vert} + {\Vert  \pv_\text{R} - \pv_{\ell} \Vert} + B}{c},
\end{align}
with $B$ indicating the clock offset (converted to meters) between the two UEs. Considering a \ac{sp} located at $\pv_{0,m}$ affecting the LOS channel and a \ac{sp} located at $\pv_{\ell,m}$ affecting the $\ell$-th RIS channel, the delay of multipath TX-SP-RX $\tau_{0,m}$ and TX-SP-RIS-RX $\tau_{\ell, m}$ can be expressed as 
\begin{align}
    \tau_{0,m} & = \frac{d_{\text{T},m} + d_{\text{R},m} + B}{c} 
    \\
    & = \frac{\Vert \pv_\text{T} - \pv_{0, m} \Vert + \Vert \pv_\text{R} - \pv_{0, m} \Vert + B}{c},\notag
    \\
    \tau_{\ell,m} & = \frac{d_{\text{T},m} + d_{\ell,m} + d_{\text{R},\ell} + B}{c} 
    \\
    & = \frac{\Vert \pv_\text{T} - \pv_{\ell, m} \Vert + \Vert \pv_\ell - \pv_{\ell, m} \Vert + \Vert \pv_\text{R} - \pv_\ell \Vert + B}{c}.\notag
\end{align}
The delay of the TX-SP-RIS-RX multipath can also be defined similarly\footnote{{Note that the single-bounce MPCs (TX-SP-RX) can be harnessed in MIMO and MISO systems~\cite{mendrzik2018harnessing, chen2022doppler}. In the SISO system, the delay estimation of the multipath is insufficient to obtain the position of the incidence point, and hence, no extra information will be provided for the UE location. By considering double- or multi-bounce MPCs (e.g., TX-SP-RIS-RX path) in RIS-aided SISO positioning, the performance might be improved. However, it is shown that the improvement is limited due to the severe pathloss, and the detection/association problems are challenging due to the multipath observation of each object via both single-bounce and double-bounce paths~\cite{kim2022ris_Double}. As a consequence, we only include MPCs in channel modeling and simulations, but the exploitation of MPCs for positioning is not considered.}}. {Considering the high pathloss of the RIS channel and the complexity of problem formulation, multi-RIS-involved paths (e.g., TX-RIS1-RIS2-RX path) are not considered throughout this work.}

The matrix $\Am_\ell(\psiv) \in \mathbb{C}^{K\times G}$ captures the effect of the $\ell$-th RIS's phase modulation with each element expressed as
\begin{align}
    [\Am_\ell(\psiv)]_{k, g} & = a_g(\psiv) = \av(\varphiv_{\text{D}})^\top \Omegam_{\ell, g} \av(\varphiv_{\text{A}})
    \label{eq_RIS_equivalent_gain_matrix}
    \\
    & = \omegav_{\ell, g}^\top (\av(\varphiv_{\text{D}}) \odot \av(\varphiv_{\text{A}})),\notag
\end{align}
where $\Omegam_{\ell, g} = \text{diag}(\omegav_{\ell, g}) \in \mathbb{C}^{N \times N}$ is a diagonal matrix and $\omegav_{\ell, g} = [\omega_{\ell, g, 1}, \ldots, \omega_{\ell, g, N}]$ ($|\omega_{\ell, g, n}|=1$) is a vector containing all the RIS element coefficients. 
The steering vectors $\av(\varphiv_\text{A})$ and $\av(\varphiv_\text{D})$ (based on the far-field assumption\footnote{\label{footnote3}{Assume a $\unit[30]{GHz}$ SIMO system with a $10\times10$ array is adopted, the Fraunhofer distance can be calculated as {$\frac{2D^2}{\lambda} = \frac{2*(\sqrt{2}*0.05)^2}{0.01} = \unit[1]{m}$}. Based on~\cite{cui2022near}, the near-field condition for a SISO system with a $10\times 10$ RIS is $\frac{d_1 d_2}{d_1 + d_2} < \frac{2D^2}{\lambda}$ with $d_1$ and $d_2$ as the distances from the RIS to TX and RX, respectively. This indicates that the far-field condition is valid if $d_1>\unit[2]{m}$ and $d_2>\unit[2]{m}$. This work considers TX and RX positions satisfying the far-field condition, and leaves near-field for future work.}}) can be expressed as
\begin{align}
    \av(\varphiv) & = 
    e^{j\frac{2\pi f_c}{c} \Zm^\top \tv(\boldsymbol{\varphi})},
    \label{eq_av_steer}
\end{align}
with $\Zm = [\zv_1, \ldots, \zv_n] \in \mathbb{R}^{3\times N}$ containing the positions of all the RIS elements {(in the local coordinate system of the RIS)}, and $\tv(\boldsymbol{\varphi})$ can be obtained from~\eqref{eq:tv_T} and~\eqref{eq:tv_R}.
{The \ac{aoa} and \ac{aod} of the $l$-th RIS path are denoted as $\psiv_\ell = [\varphiv_{\text{A},\ell}^\top, \varphiv_{\text{D},\ell}^\top]^\top$, defined in~\eqref{eq:aoa} and~\eqref{eq:aod}, while the multipath \ac{aoa}/\ac{aod} vector $\psiv_{\ell, m}$ can be defined similarly based on the position of the incidence point $\pv_{\ell, m}$.}


{
\subsection{Spatial Frequency in the RIS Channel}
In a RIS-aided positioning system, if \acp{ris} are used as reflecting anchors for \ac{tdoa}-based localization (e.g., two-step positioning in~\cite{dardari2021nlos}), at least 6 \acp{ris} are needed for 3D {sidelink} positioning due to the unknown \ac{tx} and \ac{rx} {states}. {However, by exploiting the angular information at \ac{ris}, the number of the required RIS anchors can be reduced to 2, which is the scenario considered in this work}.
In most of existing \ac{ris}-aided localization works, the state of \ac{bs} is known, and hence either $\phiv_\text{D}$ or $\phiv_\text{A}$ is known (e.g., in~\cite{keykhosravi2022ris}). When full-duplex transceivers are available, such as in self-localization or monostatic sensing scenarios~\cite{kim2022ris}, $\phiv_\text{D}$ and $\phiv_\text{A}$ are identical, and hence the channel parameters estimation of the \ac{ris} path can provide the direction of the \ac{ue} directly in most existing works.

In this sidelink positioning scenario, however, since the \ac{aod} and \ac{aoa} are both unknown, the positioning task is more challenging due to the coupling of \ac{aoa} and \ac{aod} as shown in~\eqref{eq_RIS_equivalent_gain_matrix}.}
To facilitate the positioning task, we define a steering vector~\cite{lu2022joint}
\begin{equation}
    \av_\text{R}(\varphiv_\text{D}, \varphiv_\text{A}) = \av(\varphiv_\text{D}) \odot \av(\varphiv_\text{A}) = e^{j\frac{2\pi f_c}{c} \Zm^\top \tv_\text{R}(\boldsymbol{\varphi}_\text{D}, \boldsymbol{\varphi}_\text{A})},
\end{equation}
where
\begin{equation}
    \tv_\text{R}(\boldsymbol{\varphi}_\text{D}, \boldsymbol{\varphi}_\text{A}) = \tv(\boldsymbol{\varphi}_\text{D}) + \tv(\boldsymbol{\varphi}_\text{A}).
    \label{eq_coupled_aoa_aod}
\end{equation}
Note that the first row of the matrix $\Zm$ contains all zeros (RIS elements are located on the local Y-Z plane), meaning the first element of the vector $\tv_\text{R}$ cannot be estimated.\footnote{{In the scenarios where RIS is not planar (e.g., a cylinder array), all the three elements in $\tv_\text{R}$ can be estimated directly, which is not discussed in this work.}} By further defining $\xi$ and $\zeta$ as the second and third entry of the vector $\tv_\text{R}(\boldsymbol{\varphi}_\text{D}, \boldsymbol{\varphi}_\text{A})$ given by
\begin{align}
    \label{eq:RIS_aoa_xi}
    \xi & = \sin(\phi_{\text{A}})\cos(\theta_{\text{A}}) + \sin(\phi_{\text{D}})\cos(\theta_{\text{D}}) 
    ,
    \\ 
    \zeta & = \sin(\theta_{\text{A}}) + \sin(\theta_{\text{D}}),
\end{align}
and the matrix $\Am_\ell(\psiv)$ in~\eqref{eq_RIS_equivalent_gain_matrix} can also be expressed using spatial frequencies $\xi$ and $\zeta$ as
\begin{equation}
    [\Am_\ell(\xi, \zeta)]_{g,k} = \omegav_{\ell, g}^\top \av_\text{R}(\xi, \zeta) = \omegav_{\ell, g}^\top e^{j\frac{2\pi f_c}{c} \Zm^\top [0, \xi, \zeta]^\top}.
    \label{eq:RIS_equivalent_gain_xi_zeta}
\end{equation}
These spatial frequencies (i.e., $\xi$, $\zeta$) become channel parameters to be estimated at the \ac{ris} channel, instead of the \ac{aoa} and \ac{aod}. {We note from \eqref{eq:RIS_aoa_xi} that inherent ambiguities exist in estimating \ac{aod} and \ac{aoa} from the spatial frequencies $\xi$ and $\zeta$ since many different \ac{aod}-\ac{aoa} pairs can be mapped to a single $\xi$-$\zeta$ pair, making absolute sidelink positioning task challenging.}

\subsection{Problem Statement}
\label{sec:problem_statement}
{We first define a channel parameter vector as $\etav_{\text{All}} = [\etav^\top, \etav_{\text{MP}}^\top]^\top$, where $\etav = [\etav_{0}, \etav_{1}, \ldots, \etav_{L}]\in \mathbb{R}^{5L+3}$ contains all the parameters related to LOS and RIS channels with $\etav_{0} = [\tau_0, \alpha_0, \beta_0]^\top$ and $\etav_{\ell} = [\xi_\ell, \zeta_\ell, \tau_\ell, \alpha_\ell, \beta_\ell]^\top$, and $\etav_{\text{MP}}$ contains all the multipath-related channel parameters.} In the vector $\etav_\text{All}$, $\alpha$ and $\beta$ are the amplitude and phase of the complex channel gain (i.e., $\rho=\alpha e^{-j \beta}$), $\tau$ is the delay, $\xi$ and $\zeta$ are spatial frequencies. Since the \acp{mpc} do not contribute to positioning and we do not exploit the mapping functionality in this work, we focus on the analysis of $\etav$, and the corresponding state vector $\sv = [\pv_\text{T}^\top, \pv_\text{R}^\top, B, \alpha_0, \beta_0, \ldots, \alpha_L, \beta_L]^\top \in \mathbb{R}^{3L+7}$ containing the 3D positions of two UEs, a clock offset $B$, and the complex channel gains. 
To assist analysis and algorithm development, we further define a nuisance-free channel parameter vector $\etav_\text{N} = [\eta_{\text{N},0}, \etav_{\text{N},1}, \ldots, \etav_{\text{N},L}]^\top = [\tau_0, \tau_1, \xi_1, \zeta_1, \ldots, \tau_L, \xi_L, \zeta_L]^\top$ and a nuisance-free state vector $\sv_\text{N} = [\pv_\text{T}^\top, \pv_\text{R}^\top, B]^\top \in \mathbb{R}^{7}$. {The vector $\sv_\text{N}$ may have a posterior density function $\text{Prob}(\sv)$ to optimize RIS profile for a better positioning performance, as will be discussed in Section~\ref{sec:RIS_Profile_Design}.}


Based on the definitions above, we are able to formulate the sidelink positioning problem as the estimation of the geometric channel parameter vector $\etav$ from the observed signal $\Ym$, and then calculate the state vector $\sv$ based on $\etav$ (detailed in Section~\ref{sec:channel_estimation} and~\ref{sec:positioning_algorithm}). 
To make sure the number of channel parameters (i.e., $\etav$ with $5L + 3$ elements) is larger than the number of state parameters (i.e., $\sv$ with $3L+7$ elements), {the minimum number of RISs needed is $L=2$ for the \ac{los}-available. For the \ac{los}-blockage scenario, at least $L=3$ RISs are needed.} Since the positioning task can be performed by a one-way positioning pilot signal transmission, the problem formulation can be extended to multiple UEs, {as will be discussed in Section~\ref{sec_extension_more_UEs}}. 
{Considering sidelink positioning requires multiple RISs as positioning anchors, the calibration of RISs is needed, and possible solutions can be found in~\cite{ghazalian2022bi, ghazalian2022joint, lu2022joint,zheng2023jrcup}, which will not be discussed in this work.}

\section{Lower Bound Analysis}
\label{sec_performance_analysis}
In this section, we derive the \acp{crb} for the estimation of the channel parameter vector $\etav$ and state vector $\sv$, based on the \ac{fim} analysis.~\footnote{{In this work, we only consider the \ac{fim} provided by the data (i.e., received signal) as $\bm{\mathcal{I}} = \bm{\mathcal{I}}_\text{D}$ for performance analysis and positioning algorithm design. If the position prior information is available, the corresponding \ac{fim} $\bm{\mathcal{I}}_\text{P}$ will only be used for \ac{ris} profile design in Section~\ref{sec_methodology}. However, $\bm{\mathcal{I}}_\text{P}$ can also be easily included as $\bm{\mathcal{I}} = \bm{\mathcal{I}}_\text{D} + \bm{\mathcal{I}}_\text{P}$, for performance analysis (e.g., Bayesian \ac{crb}) or estimator development (e.g., maximum a posteriori estimator)}.}

\subsection{CRB of Channel Parameter Estimation}
Based on the defined channel parameter vector $\etav_\text{All}$, state vector $\sv$, and the signal model in~\eqref{eq:signal_model},~\eqref{eq:channel_LOS_MP},~\eqref{eq:channel_RIS_MP}, the \ac{fim} of channel parameter estimation can be obtained as~\cite{kay1993fundamentals} (Sec. 3)
\begin{align}
    \bm{\mathcal{I}}({\boldsymbol\eta_\text{All}}) & 
    = \frac{2}{\sigma_n^2} \sum^K_{k=1}\sum^{G}_{g=1}\mathrm{Re}\left\{
    \left(\frac{\partial{\mu}_{k,g}}{\partial{\boldsymbol\eta_{\text{All}}}}\right)^{\mathsf{H}} 
    \left(\frac{\partial{\mu}_{k,g}}{\partial{\boldsymbol\eta_{\text{All}}}}\right)\right\}.
    \label{eq:FIM_measurement}
\end{align}
Here, $\mu_{k,g} = [\Ym_{\text{U}}]_{k,g} + [\Ym_{\text{R}}]_{k,g}$ is the noise-free observation of the received signal. {Note that for fixed system parameters and pilot signals, the $\bm{\mathcal{I}}({\boldsymbol\eta_\text{All}})$ depends on the system geometry (i.e., the position of \ac{tx} and \ac{rx}, the position/orientation of \acp{ris}, and \ac{ris} profiles}.

{In contrast to a MIMO system where multipath components can help in localization, \acp{sp} in the SISO or SIMO/MISO scenarios (e.g., multipath in the LOS and RIS channels of this work) do not affect the localizability of the problem but may degrade the performance of the algorithm (i.e., resulting in unresolvable paths). More discussions on the bounds considering \ac{nlos} paths can be found in~\cite{ge2023analysis}. 
In this work, we focus on the non-multipath channel parameters $\etav$, whose \ac{fim} can be easily obtained as the equivalent \ac{fim}~\cite{chen2022tutorial} of the top-left submatrix as 
$\bm{\mathcal{I}}({\boldsymbol\eta}) \in \mathbb{C}^{(5L+3)\times (5L+3)}$.}
We can further define delay error bound (DEB) and spatial error bounds (SEBs) for $\tau_\ell, \xi_\ell, \zeta_\ell$ as
\begin{align}
\mathrm{EB}_{\tau_\ell} & = \sqrt{[\bm{\mathcal{I}}({\boldsymbol\eta})^{-1}]_{1+5(\ell-1), 1+5(\ell-1)}}\ , \ \ \ (\ell \ge 0),
\label{eq:DEB}\\
\mathrm{EB}_{\xi_\ell} & = \sqrt{[\bm{\mathcal{I}}({\boldsymbol\eta})^{-1}]_{5\ell-1, 5\ell-1}}\ ,
 \ \ \ (\ell>0),
\label{eq:SEB1}\\
\mathrm{EB}_{\zeta_\ell} & = \sqrt{[\bm{\mathcal{I}}({\boldsymbol\eta})^{-1}]_{5\ell, 5\ell}}\ , \ \ \ (\ell>0).
\label{eq:SEB2}
\end{align}


\subsection{CRB for 3D Sidelink Positioning}
Based on $\bm{\mathcal{I}}({\boldsymbol\eta})$, the CRB of the state parameters $\sv$ can be obtained as 
\begin{equation} \label{eq_crb_sidelink}
    \mathrm{CRB} \triangleq \left[\bm{\mathcal{I}}(\sv)\right]^{-1} = \left[\Jm_\mathrm{S} \bm{\mathcal{I}}({\etav}) \Jm_\mathrm{S}^\top\right]^{-1},
\end{equation}
where $\Jm_\mathrm{S} \triangleq \frac{\partial {\etav}}{\partial \sv} \in \mathbb{R}^{(3L+7)\times(5L+3)}$ is the Jacobian matrix using a denominator-layout notation from the channel parameter vector $\etav$ to the state vector $\sv$.
We can further define the position error bounds (PEBs), and clock offset error bound (CEB) as
\begin{align} 
\mathrm{PEB_\text{T}} & = \sqrt{\trace([\bm{\mathcal{I}}({\sv})^{-1}]_{1:3, 1:3})}
\label{eq:PEBT},\\  
\mathrm{PEB_\text{R}} & = \sqrt{\trace([\bm{\mathcal{I}}({\sv})^{-1}]_{4:6, 4:6})}
\label{eq:PEBR},\\
\mathrm{CEB} & = \sqrt{([\bm{\mathcal{I}}({\sv})^{-1}]_{7, 7})}
\label{eq:CEB}.
\end{align}
{The derived CRB can be used to assist RIS profile design (see Section~\ref{sec:RIS_Profile_Design}), select weight coefficients in positioning algorithm (see Section~\ref{sec:positioning_algorithm}), and evaluate the proposed positioning algorithm as well as RIS profile design performance (see Section~\ref{sec_simulation_results}).}

\section{Methodology}
\label{sec_methodology}
In this section, we describe RIS profile design strategy, codebooks for the scenarios with and without prior information, channel parameter estimation algorithms, and positioning algorithms.\footnote{Note that the effect of multipath is not considered when designing the RIS profile and positioning algorithms. {For the architectures/protocols of the RIS-aided sidelink positioning systems and the coordination between the involved devices (e.g., RISs and UEs), please check~\cite{chen2023riss} for more details.}} 

\subsection{RIS Profile Design}
\label{sec:RIS_Profile_Design}
{In the positioning phase, we assume the RIS profiles (or codebooks) are always known at the UE side via RIS-aided positioning protocols~\cite{chen2023riss}.
To assist channel parameter estimation, we adopt \textit{time-orthogonal RIS profiles} to differentiate independent RIS paths from the others~\cite{keykhosravi2022ris}. We first divide the total transmission $G$ into $\Gamma \ge L+1$ blocks (each block with $\tilde G = G/\Gamma$ OFDM symbols) and define a matrix $\Bm \in \mathbb{C}^{\Gamma \times (L+1)}$  containing orthogonal columns (e.g., from a DFT matrix) as~\cite{dardari2021nlos, keykhosravi2021semi}
\begin{equation}
    \Bm = [\bv_0, \bv_1, \ldots, \bv_{L}],\ \ \ \ \text{s.t.}\ \ \Bm^\herm\Bm = \mathbf{I}_{(L+1)\times(L+1)},
\end{equation}
where each element inside $\Bm$ has a unit amplitude (i.e., $|[\Bm]_{i,j}|=1$). By selecting $\omegav_{\ell,\tilde g} \in \mathbb{C}^{N}$ for $1\le \ell \le L$, and $1\le \tilde g\le \tilde G$, the rest of the RIS profiles can be obtained as
\begin{equation}
\omegav_{\ell, (i-1)\tilde G+\tilde g} = b_{\ell,i}^*\omegav_{\ell, \tilde g},\ (i = 1,\ldots, \Gamma),
\label{eq:orthogonal_profiles}
\end{equation}
where $b_{\ell, i}$ is the $i$-th element of the vector $\bv_{\ell}$. We further define the received signal for the $i$-th block as $\Ym^{(i)}$ ($i = 1, \ldots, \Gamma$), and the LOS path and all the RIS paths can be separated as
\begin{align} \label{eq_ytilde}
    \tilde \Ym_\ell & = \frac{1}{\Gamma}\sum_{i=1}^\Gamma b_{ \ell,i}\Ym^{(i)} 
    = \Ym^{(1)}_\ell + \tilde \Nm,
\end{align}
where $\tilde \Ym_\ell \in \mathbb{R}^{K\times \frac{G}{\Gamma}}$ has a smaller size than the received signal block $\Ym \in \mathbb{R}^{K\times G}$ defined in~\eqref{eq:signal_model}, and $\tilde \Nm \in \mathbb{R}^{K \times \tilde G}$ with each element $n_{k,\tilde g} \sim \mathcal{CN}(0, \sigma_n^2/\Gamma)$. 

In the following RIS profile design, we will only discuss the design of the first block, and the rest of the blocks can be obtained based on~\eqref{eq:orthogonal_profiles} to form orthogonal profiles that assist channel parameter estimation.
Without any prior information on the UE positions, \textit{random codebooks} are adopted. If the prior information is available, \textit{directional} or \textit{directional and derivative codebooks} can be used, as detailed below.}

\subsubsection{Random Codebook}
Without any UE position prior information, a random codebook can be adopted. In this case, each element in the coefficients vector of the $\ell$-th RIS $\omegav_{\ell, g}$ is chosen with unit amplitude and random phase following $\angle \omega_{{\ell, g, n}} \sim \mathcal{U}[0, 2\pi)$.

\subsubsection{Directional Codebook}
{Assume the position prior information of both \acp{ue} is available (e.g., from previous estimations or based on statistical information), RIS profiles can be designed to improve positioning performance~\cite{keskin2022optimal, fascista2022ris}.} A directional (DIR) codebook is one of the simplest codebooks with the main idea of maximizing the received signal strength of the receiver, given the prior position information of two UEs. 
By dropping the time index $g$, for the UEs located at $\pv_\text{T}$ and $\pv_\text{R}$, the optimal RIS profile that maximizes the received energy can be obtained based on~\eqref{eq_RIS_equivalent_gain_matrix} and~\eqref{eq_av_steer} as
\begin{equation}
    \omegav^{(1)}_\ell = \av^*(\varphiv_\ell) = e^{-j\frac{2\pi f_c}{c}\Zm^\top (\tv_{\text{T},\ell} + \tv_{\text{R},\ell})}.
    \label{eq:RIS_profile_sum_beam}
\end{equation}
Here, $\tv_{\text{T},\ell}$ and $\tv_{\text{R},\ell}$ can be obtained based on~\eqref{eq:tv_T} and~\eqref{eq:tv_R}, which are the direction vectors obtained from
$\pv_{\text{T}}$ and $\pv_{\text{R}}$, respectively. In real scenarios, however, we cannot know the true location of both UEs, and the prior information may appear in the form of certain distributions 
{(e.g., the posterior distribution of the nuisance-free state vector $\text{Prob(\sv)}$ with $\sv_\text{N} \sim \mathcal{N}(\bar \sv_\text{N}, \bar \Sigmam_{\sv_\text{N}}$).}
In this case, we can sample two candidate positions $\pv_{\text{T},\tilde g}$ and $\pv_{\text{R},\tilde g}$ ($1\le \tilde g \le \tilde G$) for $\tilde G$ times based on $\text{Prob}(\sv_\text{N})$. The DIR codebook of the $\ell$-th RIS (for the first block of transmissions) can be obtained as 
\begin{equation}
\Xim_\ell^{\text{DIR}} = [\omegav_{\ell,1}^{(1)}, \ldots, \omegav_{\ell,\tilde G}^{(1)}]  \in \mathbb{C}^{N\times \tilde G},
\label{eq_DIR_codebook}
\end{equation}
with each column $\omegav_{\ell,\tilde g}^{(1)}$ corresponding to the DIR beam sampled UE positions $\pv_{\text{T},\tilde g}$ and $\pv_{\text{R},\tilde g}$ based on~\eqref{eq:RIS_profile_sum_beam}. The RIS profiles for the rest $(\Gamma-1)$ blocks can be obtained based on~\eqref{eq:orthogonal_profiles}.


\subsubsection{Directional and Derivative Codebook}
As has been shown in previous works~\cite{fascista2022ris, keskin2022optimal}, maximizing the received signal strength {at the targeted direction} does not {necessarily lead to} an optimal positioning performance. {Hence, the SNR-maximizing directional codebook can be suboptimal from the perspective of positioning.} For given $\xi$ and $\zeta$ (computed from the positions $\pv_{\text{T}}$ and $\pv_{\text{R}}$), 
the optimal RIS phase profiles {(in the sense of minimizing the PEBs in \eqref{eq:PEBT} and \eqref{eq:PEBR}} should lie in the subspace spanned by the following vectors {\cite[Section~IV-A]{crb_mimo_radar_2008},\cite[Prop.~1]{keskin2022optimal}}:
\begin{align}
    \omegav^{(1)} &= \avr^*(\xi, \zeta) 
    = e^{-j\frac{2\pi f_c}{c}\Zm^\top [0,  \xi, \zeta]^\top}, 
    \label{eq:sum_diff_beam_1}
    \\
    \omegav^{(2)} &= \frac{\partial \avr^*( \xi,  \zeta) }{\partial \xi}
    = \omegav^{(1)} \odot ({-j\frac{2\pi f_c}{c}\Zm^\top [0, 1, 0]^\top}),
    \label{eq:sum_diff_beam_2}
    \\
    \omegav^{(3)} &= \frac{\partial \avr^*( \xi,  \zeta) }{\partial \zeta}
    = \omegav^{(1)} \odot ({-j\frac{2\pi f_c}{c}\Zm^\top [0, 0, 1]^\top}),
    \label{eq:sum_diff_beam_3}
\end{align}
where the first RIS profile $\omegav^{(1)}$ is identical to the {\textit{directional}} (DIR) beam defined in~\eqref{eq:RIS_profile_sum_beam}, and $\omegav^{(2)}$, $\omegav^{(3)}$ are the so-called \textit{derivative (DER)} beams. Since the elements in $\omegav^{(2)}$ and $\omegav^{(3)}$ do not have unit amplitude, gradient projection~\cite{tranter2017fast} is adopted to find the closest unit-amplitude profiles to $\omegav^{(2)}$ and $\omegav^{(3)}$. 
{The intuition behind the derivative beams is to induce large amplitude changes in response to small perturbations in spatial frequency in the local neighborhood of the true value (see \cite[Fig. 2]{keskin2022optimal}). This enables accurate estimation of spatial frequency, along with the high SNR provided by the directional beam.}
Similar to the formulation of the DIR codebook in~\eqref{eq_DIR_codebook}, we can sample TX and RX UE positions for $\tilde G/3$ times based on their distribution. The DIR+DER codebook of the $\ell$-th RIS $\Xim_\ell \in \mathbb{C}^{N\times \tilde G}$ can be formulated as
\begin{equation}
\Xim_\ell = [\omegav_{\ell,1}^{(1)}, \omegav_{\ell,1}^{(2)}, \omegav_{\ell,1}^{(3)}, \ldots, \omegav_{\ell,\tilde G/3}^{(1)}, \omegav_{\ell,\tilde G/3}^{(2)}, \omegav_{\ell,\tilde G/3}^{(3)}]  .
\label{eq_DER_codebook}
\end{equation}
Similarly, the RIS profiles for the rest $(\Gamma-1)$ blocks can be obtained based on~\eqref{eq:orthogonal_profiles}.


\subsubsection{Power Control of the DIR+DER Codebook with Prior Information}
To further improve the positioning performance, power control can be adopted for the implemented codebook by using different transmit powers for each beam in the codebook. 
More specifically, the optimization problem can be formulated (take RX UE for example) as minimizing the expectation of the {squared} RX PEB (defined in~\eqref{eq:PEBR}) given by
\begin{align}
\min_{\bm\delta \in \mathbb{R}^G} \int \text{Prob}(\sv) 
& \mathrm{PEB^2_\text{R}}(\sv, \Xim_1, \ldots,  \Xim_L|\deltav) \mathrm{d} \sv,     \label{eq_opt_formulation_1}
\\
& \mathrm{s.t.} ~~  \Vert \deltav \Vert ^2 = G ~,\notag
\end{align}
where $\deltav$ is the power control vector defined in~\eqref{eq_transmission_matrix}, and $\text{Prob}(\sv)$ is the posterior distribution of the state vector $\sv$. Considering the high complexity of solving the problem~\eqref{eq_opt_formulation_1}, especially when the integral operation is involved, we can further simplify the problem formulation as
\begin{align}
  \min_{\bm\delta \in \mathbb{R}^G} 
  \frac{3}{\tilde G}\sum_{\tilde g = 1}^{\tilde G/3} & \mathrm{PEB^2_\text{R}}(\pv_{\text{T}, \tilde g}, \pv_{\text{R}, \tilde g}, \Xim_1, \ldots,  \Xim_L|\deltav),  
  \label{eq_opt_formulation_2}
  \\
  ~~ & \mathrm{s.t.} ~~  \Vert \deltav \Vert ^2 = G ~,\notag
\end{align}
where $\pv_{\text{T}, \tilde g}, \pv_{\text{R}, \tilde g}$ are sampled based on the prior information. \eqref{eq_opt_formulation_2} can be {equivalently reformulated as a convex problem
(see Appendix~\ref{app_conv_form}), 
which can be} solved using convex optimization {tools to} provide optimal positioning performance for a given codebook~\cite{fascista2022ris, keskin2022optimal}. Based on the insights from simulation results, optimal performance can be achieved when DER beams $\omegav^{(2)}$ and $\omegav^{(3)}$ are assigned with the same amount of power. In order to further relieve the computational burden, we propose to use the same power control coefficient $\frac{\sqrt{3}}{\sqrt{1+2\gamma_\text{P}^2}}$ for all the DIR beams (i.e., $\omegav^{(1)}$), and the same coefficient $\frac{\sqrt{3}\gamma_\text{P}}{\sqrt{1+2\gamma_\text{P}^2}}$ for all the DER beams (i.e., $\omegav^{(2)}$, $\omegav^{(3)}$), where $\gamma_\text{P}$ is the ratio between of the DER beam power and DIR beam power (only DIR beams are kept when $\gamma_\text{P}=0$). The optimization problem in~\eqref{eq_opt_formulation_2} can be simplified as
\begin{align}
  \min_{\gamma_\text{P} \in \mathbb{R}} 
  \sum_{\tilde g = 1}^{\tilde G/3} \mathrm{PEB^2_\text{R}}(\pv_{\text{T}, \tilde g}, \pv_{\text{R}, \tilde g}, \Xim_1, \ldots,  \Xim_L|\gamma_\text{P}),   ~~ \gamma_\text{P} \ge 0 ~.
  \label{eq_opt_formulation_3}
\end{align}

\subsection{Channel Parameter Estimation Algorithm}
\label{sec:channel_estimation}
Once the RIS profiles are designed, the system can send positioning pilot signals and perform positioning algorithms. Here, we describe a two-stage positioning algorithm, including a channel parameter extraction step and a positioning step. For each stage, a coarse estimation algorithm and a refined \ac{mle} are developed for different performance and complexity tradeoffs. 

\subsubsection{Low-complexity Channel Parameters Estimator}
By implementing the orthogonal RIS profile as described in Section~\ref{sec:RIS_Profile_Design}, the uncontrolled path and each RIS path can be well-separated.
For the LOS path observation $\tilde \Ym_0$ from \eqref{eq_ytilde}, we first obtain the estimated channel elements and sum across all the $G$ transmissions as
\begin{equation}
    \underbrace{\hv_0}_{{\hv_0\in \mathbb{R}^K}} = \sum_{g=1}^{\tilde G} [\tilde\Ym_0]_{:,g}
\odot \xv^*.
\label{eq:coarse_los_channel}
\end{equation}
The delay of the LOS path $\hat \tau_0$ can be estimated based on~\eqref{eq:signal_model}, \eqref{eq:channel_LOS_MP} and \eqref{eq:coarse_los_channel} as
\begin{equation}
    \hat \tau_0 = \arg\max_{\tau} |\dv^\hermit(\tau) \hv_0|,
    \label{eq_coarse_los_delay}
\end{equation}
where $\dv(\tau)$ is defined in~\eqref{eq:delay_matrix}, and~\eqref{eq_coarse_los_delay} can be solved using an $N_\text{F}$ point \ac{dft}~\cite{fascista2022ris}.
For the observation of the $l$-th RIS path $\tilde \Ym_\ell$, since RIS profiles are different from one transmission to another, we need to modify~\eqref{eq:coarse_los_channel} and~\eqref{eq_coarse_los_delay} as
\begin{align}
    \underbrace{\hat \Hm_\ell}_{\hat \Hm_\ell\in \mathbb{R}^{K\times \tilde G}} & = \tilde\Ym_\ell \odot (\xv^* \mathbf{1}_{\tilde G}^\top), \label{eq:coarse_ris_H} \\
    \hat \tau_\ell & = \arg \max_{\tau} \Vert \dv^\hermit(\tau) \hat \Hm_\ell   \Vert. \label{eq_coarse_delay_RIS}
\end{align}
Once the delay of the $\ell$-th RIS path $\tau_\ell$ has been obtained, the estimation of spatial frequencies $\hat \xi_\ell$ and $\hat \zeta_\ell$ can be formulated as
\begin{equation}
    \label{eq:coarse_ris_xizeta}
    [\hat \xi_\ell, \hat \zeta_\ell] = \argmin_{\xi, \zeta} \sum_{g,k} |\omegav_{\ell, g}^\top e^{\frac{j 2\pi}{\lambda_c}\Zm^\top [0, \xi, \zeta]^\top}d_k(\hat \tau_\ell) x_k \tilde y_{k,g}^*|,
\end{equation}
where $\tilde y_{k,g}$ is the element of the matrix $\tilde \Ym$, and the problem can be solved via a 2D search.


\subsubsection{MLE for Channel Parameter Estimation}
From the low-complexity channel parameters estimator, we can estimate the nuisance-free channel parameter vector $\hat \etav_{\text{N}, \ell}$ ($\ell = 0, 1, \ldots, L$). The MLE aims to find the optimal channel parameters as 
\begin{equation}
    [\hat \rho_\ell, \hat \etav_{\text{N},\ell}] = \argmin_{\rho_\ell, \boldsymbol{\eta}_{\text{N},\ell}}
    \Vert \tilde\muv_\ell - \rho_{\ell}\muv_\ell(\etav_{\text{N},\ell})\Vert,
    \label{eq_mle_channel_estimation_0}
\end{equation}
where $\tilde\muv_\ell = \text{vec}(\tilde \Ym_\ell)$, $\muv^\herm(\etav_{\text{N},0}) = \text{vec}(\Dm(\tau_0)\odot\Xm)$, and $\muv^\herm(\etav_{\text{N},\ell}) = \text{vec}(\Dm(\tau_\ell)\odot \Am_{\ell}(\xi_\ell, \zeta_\ell)\odot\Xm)$ ($\ell\ge 1$) that can be obtained from~\eqref{eq:delay_matrix} and~\eqref{eq:RIS_equivalent_gain_xi_zeta}. 
Since the channel gain $\rho_\ell$ is a complex constant, by letting $\partial \Vert \tilde\muv_\ell - \rho_{\ell}\muv_\ell(\etav_{\text{N},\ell}) \Vert^2/\partial \rho_{\ell} = 0$, we can obtain the channel gain as $\hat \rho_\ell = \frac{\muv^\herm(\etav_{\text{N},\ell})\tilde\muv}{\Vert \muv(\etav_{\text{N},\ell}) \Vert^2}$. 
And hence, the \ac{mle} can be formulated from~\eqref{eq_mle_channel_estimation_0} with nuisance-free channel parameters only as
\begin{equation}
    \hat \etav_{\text{N},\ell} = \argmin_{\etav_{\text{N},\ell}}\left\Vert \tilde \muv_{\ell} - \frac{\muv^\herm(\etav_{\text{N},\ell})\tilde\muv}{\Vert \muv(\etav_{\text{N},\ell}) \Vert^2}\muv_{\ell}(\etav_{\text{N},\ell})\right\Vert.
    \label{eq:mle_channel_estimation}
\end{equation}

\begin{algorithm}[t]
\small
\caption{Channel Parameter Estimation}
\label{alg_channel_estimation}
\begin{algorithmic}[1]
\State \textbf{--- \textit{Coarse Estimation} ---}
\State Input: $\tilde \Ym_0$, $\tilde \Ym_\ell,\ \ell=1,\dots,L.$
\State Estimate $\hat{\eta}_{\text{N}, 0}=\hat{\tau}_0$ using \eqref{eq:coarse_los_channel}, \eqref{eq_coarse_los_delay}.
\For{$\ell = 1$ to $L$}
    \State Estimate $\hat{\tau}_\ell$ using \eqref{eq:coarse_ris_H} and \eqref{eq_coarse_delay_RIS}.
    \State Estimate $\hat \xi_\ell$ and $\hat \zeta_\ell$ using 2D grid search in \eqref{eq:coarse_ris_xizeta}.
    \State $\hat{\etav}_{\text{N}, \ell}$ $\leftarrow$ $[\hat{\xi}_\ell, \hat{\zeta}_\ell, \hat{\tau}_\ell]^\top$.
\EndFor
\Return $\hat{\etav}_{\text{N}, \ell}, \ell=0,\dots,L$.
\State \textbf{--- \textit{Refinement} ---}
\State Input: Coarse estimates $\hat{\etav}_{\text{N}, \ell}, \ell=0,\dots,L$.
\For{$\ell = 0$ to $L$} 
    \State Obtain the refined $\hat \etav_{\text{N},\ell}$ by solving~\eqref{eq:mle_channel_estimation} initialized with $\hat{\etav}_{\text{N}, \ell}$.
\EndFor
\Return refined $\hat{\etav}_\text{N} = [\hat{\eta}_{\text{N}, 0}, \hat{\etav}_{\text{N}, 1}, \ldots, \hat{\etav}_{\text{N}, L}]^\top$.
\end{algorithmic}
\end{algorithm}

\subsection{Positioning Algorithm}
\label{sec:positioning_algorithm}
\subsubsection{Coarse Position Estimation}
Based on the estimated channel parameter vector $\hat\etav$, we propose a 3D-search positioning algorithm to estimate 7 unknowns (i.e., $\pv_\text{T}$, $\pv_\text{R}$, and $B$). For a position candidate $\check \pv_\text{T}$ of the transmitter UE, the candidate direction vector $\check \tv_{\text{T},\ell}$ can be obtained from~\eqref{eq:tv_T}. Based on the estimated spatial frequency $\hat \xi_\ell$ and $\hat \zeta_\ell$, the candidate direction vector of the $\ell$-th RIS $\check \tv_{\text{R},\ell}$ can be calculated as 
\begin{equation}
    \label{eq_candidate_direction_vec_RX}
    \begin{split}
    \check t_{\text{R},\ell, 2} & = \hat\xi_\ell - \check t_{\text{T},\ell, 2}\ ,\\
    \check t_{\text{R},\ell, 3} & = \hat\zeta_\ell - \check t_{\text{T},\ell, 3}\ ,\\
    \check t_{\text{R},\ell, 1} & = \sqrt{1-\check t_{\text{R},\ell, 2}^2-\check t_{\text{R},\ell, 2}^2}\ .
    \end{split}
\end{equation}
Note that ambiguities exist in the estimated spatial frequencies due to $\hat \xi, \hat \zeta \in [-1, 1)$, while the true spatial frequencies $\xi, \zeta \in [-2, 2]$. This issue can be solved with prior location information to limit the searching area, or with a reduced RIS inter-element spacing (e.g., to $\lambda_c/4$ instead of $\lambda_c/2$, see~\cite{kim2022ris}).

Based on the candidate direction vector $\check \tv_{\text{R},\ell}$ ($\ell \ge 1$) and known RIS states, we are able to calculate the candidate receiver UE position $\check \pv_{\text{R}}$ by getting the closest point to both \ac{aod} direction vectors~\cite{brandstein1997closed}. Given two bearing lines $\lv_i = \pv_i+r \check \tv_{\text{G},i}$ and $\lv_j = \pv_j+r \check \tv_{\text{G},j}$ ($i,j \in {1, \ldots, L}$ and $\check \tv_{\text{G},\ell} = \Rm_\ell \check \tv_{\text{R},\ell}$), the following equations hold
\begin{align}\label{eq:a1}
    \check  \pv_{ij} - \check  \pv_{ji} & = -\check d_{ji}(\check \tv_{\text{G}, j}\times \check \tv_{\text{G}, i}),\\
    \check d_{ji} & = \frac{(\check \tv_{\text{G}, j}\times \check \tv_{\text{G}, i})(\pv_j - \pv_i)}{|\check \tv_{\text{G}, j}\times \check \tv_{\text{G}, i}|},
\end{align}
where $\check \pv_{ij}$ is the closest point on the bearing line $\lv_i$ to the bearing line $\lv_j$ denoted as
\begin{equation}
    \check \pv_{ij} = \pv_i + \check r_{ij} \check \tv_{\text{G},i}.
    \label{eq:closest_points}
\end{equation}
By using least squares, $r_{ij}$ and $r_{ji}$ can be obtained as
\begin{equation}
    \begin{bmatrix}
    \check r_{ij} \\ \check r_{ji}
    \end{bmatrix}
    = (\check \Qm^\top \check \Qm)^{-1}\check \Qm^\top [\pv_j - \pv_i -\check d_{ji}(\check \tv_{\text{G}, j}\times \check \tv_{\text{G}, i})],
\end{equation}
with $\check \Qm = [\check \tv_{\text{G},i}, \check \tv_{\text{G},j}]$, and the candidate receiver UE position can be obtained as 
\begin{equation}\label{eq:candidate_UE}
    \check \pv_\text{R} = \sum_{{i< j}} w_{ij}\check\pv_{ij}, \ \  (\sum_{{i< j}} w_{ij}=1),
\end{equation}
where $i, j\in \{1,\ldots, L\}$ and $w_{ij}$ is the weight coefficient that can be chosen {equally as $\frac{2}{L(L-1)}$ if no prior information is provided or based on the quality of the channel parameters estimation (e.g., SNR or PEB using prior information). Since this work is an early-stage work for sidelink positioning, we only perform simulation results (see Section~\ref{sec_weighting_coefficients_3RIS}) to illustrate the effect of different weight coefficients on estimation results. The optimization of weight selection is left for future work.}


Finally, we can obtain the estimated clock offset $\check B$ as
\begin{equation}
\label{eq_coarse_position_checkB}
    \check B = c \check \tau_0 - \Vert \check \pv_{\text{R}} - \check \pv_{\text{T}} \Vert,
\end{equation}
and the cost function can be formulated as 
\begin{equation}\label{eq:coarse_position_costFun}
    J(\check \pv_\text{T}) = \sum_{\ell=1}^L w_\ell 
    |\check B + \Vert \check \pv_{\text{T}} - \check \pv_\ell \Vert + \Vert \check \pv_{\text{R}} - \check \pv_\ell \Vert - c \check \tau_\ell|,
\end{equation}
with $w_\ell$ as the weight coefficient {depending on the quality of the delay measurement, which can be set as $1$ by default. Note that the weights mainly affect coarse estimation, but their impacts on the refined results are limited, as will be shown in Section~\ref{sec_weighting_coefficients_3RIS}}.
Among all the transmitter UE position candidates, the one with the lowest cost will be the estimated position as
\begin{equation}\label{eq:coarse_position_hatpv}
    \hat \pv_\text{T} = \argmin_{\check \pv_\text{T}} J(\check \pv_\text{T}),
\end{equation}
and the rest of the state parameter vector can be obtained based on~\eqref{eq_candidate_direction_vec_RX} to~\eqref{eq_coarse_position_checkB}.

\begin{remark}
{The coarse position estimation algorithm can be easily extended into the \ac{los}-blockage scenarios by estimating the clock offset from the shortest \ac{ris} path (take $\ell = 1$ for example) as
\begin{equation}
    \check B = c\check \tau_1 - \Vert \check\pv_\text{R} - \pv_\text{1} \Vert - \Vert \check\pv_\text{T} - \pv_\text{1} \Vert,
\end{equation}
and the summation in the cost function~\eqref{eq:coarse_position_costFun} starts from $\ell = 2$.}  
\end{remark}

\subsubsection{MLE for Positioning}
The MLE refinement for positioning can be formulated as
\begin{equation}
    \hat \sv_{\text{N}} = \argmin_{\sv_{\text{N}}}\ \left(\hat \etav_{\text{N}} - \etav_{\text{N}}(\sv_{\text{N}})\right)^\mathsf{T}
    \Sigmam_{\bm{\eta}_\text{N}}^{-1}\left(\hat \etav_{\text{N}} - \etav_{\text{N}}(\sv_{\text{N}})\right),
    \label{eq:mle_position_estimation}
\end{equation}
where $\Sigmam_{\eta_\text{N}} = \bm{\mathcal{I}}({\boldsymbol\eta}_\text{N})^{-1}$ is the covariance matrix of the estimated channel parameters, and the optimization problem in~\eqref{eq:mle_position_estimation} can be solved by, e.g., the trust-region method and the gradient of the cost function in~\eqref{eq:mle_position_estimation} is $-(\frac{\partial \etav_{\text{N}}(\sv_{\text{N}})}{\sv_{\text{N}}})^\mathsf{T}\bm{\Sigma_{\eta_\text{N}}}^{-1}\left(\hat \etav_{\text{N}} - \etav_{\text{N}}(\sv_{\text{N}})\right)$. 
For the scenarios covariance matrix in MLE formulation is not available, we can set $\Sigmam_{\bm{\eta}_\text{N}}=\mathbf{I}$, leading to a least squares solution. The pseudo-codes for channel parameter estimation and position estimation can be found in Algorithm~\ref{alg_channel_estimation} and Algorithm~\ref{alg_position_estimation}, respectively.


\begin{algorithm}[t]
\small
\caption{Position Estimation}
\label{alg_position_estimation}
\begin{algorithmic}[1]
\State \textbf{--- \textit{Coarse Estimation} ---}
\State Input: Channel parameters $\hat{\etav}_\text{N}$, searching area $\mathcal{A}$.
\For{candidate transmitter $\check \pv_\text{T}\in\mathcal{A}$}
    \State Compute $\check \tv_{\text{T},\ell},\ \ell=1,\dots,L$ based on~\eqref{eq:tv_T}.
    \State Compute $\check \tv_{\text{R},\ell},\ \ell=1,\dots,L$ based on~\eqref{eq_candidate_direction_vec_RX}.
    \State Obtain the candidate UE position $\check \pv_{\text{R}}$ through~\eqref{eq:a1}--\eqref{eq:candidate_UE}.
    \State Compute $\check{B}$ using~\eqref{eq_coarse_position_checkB}.
    \State Calculate the cost $J(\check \pv_\text{T})$ in~\eqref{eq:coarse_position_costFun}.
\EndFor
\State Select the optimal $\hat \pv_\text{T}$ that minimizes $J(\check \pv_\text{T})$ as~\eqref{eq:coarse_position_hatpv} and the corresponding $\hat \pv_\text{R}$ and $\hat B$.
\State 
\Return $\hat\pv_\text{T}, \hat\pv_\text{R}, \hat B$.
\State \textbf{--- \textit{Refinement} ---}
\State Input: coarse estimates $\hat\pv_\text{T}, \hat\pv_\text{R}, \hat B$.
\State Obtain refined $\hat \sv_\text{N}$ by solving~\eqref{eq:mle_position_estimation} with the initialization $\{\hat\pv_\text{T}, \hat\pv_\text{R}, \hat B\}$.
\State
\Return refined $\hat\sv_\text{N} = [\hat\pv_\text{T}^\top, \hat\pv_\text{R}^\top, \hat B]^\top$.
\end{algorithmic}
\end{algorithm}

\subsection{Complexity Analysis}
In this subsection, we perform complexity analysis on the proposed channel parameter estimation in~Section~\ref{sec:channel_estimation} and positioning algorithms in Section~\ref{sec:positioning_algorithm}. 
In channel parameter estimation, $L+1$ 1D $N_\text{F}$-point \ac{dft} are needed for delay estimation, resulting in complexity on the order of $\mathcal{O}_{1}(LN_\text{F}\log N_\text{F})$. For each of the $L$ RISs, a 2D search for spatial frequency estimation is needed, resulting in a complexity of $\mathcal{O}_2(LQ_1 Q_2 GKN)$, where $Q_1=|\mathcal{G}_\xi|$ and $Q_2=|\mathcal{G}_\zeta|$ denote the searching dimension of $\xi$ and $\zeta$, respectively. To refine the channel parameter estimation, we have $\mathcal{O}_3(LQ_3 GKN)$, where $Q_3$ is the number of iterations.
Regarding the positioning algorithm, a 3D search is needed to estimate the positions of both UEs and clock offset, giving $\mathcal{O}_4(L^2Q_4 Q_5 Q_6)$, where $L^2$ indicates the number of beam pairs to be calculated (e.g., there are $L(L-1)$ pairs of beams from $L$ RISs to obtain the candidate receiver position via intersections) and $Q_4, Q_5, Q_6$ represent the searching dimension of the position on the $x$, $y$, and $z$ axis, respectively. For the refinement via MLE, the complexity is $\mathcal{O}_5(L^2 Q_7)$, where {$L^2$} indicates the multiplication of matrices, and $Q_7$ is the number of iterations to refine the positioning results. 
In summary, the overall complexity of the positioning problem is given by
\begin{align}
& \mathcal{O}_\text{P} = \notag
\\
& \underbrace{\mathcal{O}_1(LN_\text{F}\log N_\text{F}) + \mathcal{O}_2(LQ_1 Q_2 GKN) + \mathcal{O}_4(L^2Q_4 Q_5 Q_6)}_{\text{Coarse Estimation}}\notag
\\
& + \underbrace{\mathcal{O}_3(LQ_3 GKN) + \mathcal{O}_5(L^2 Q_7)}_{\text{Refinement}}. 
\label{eq_complexity_analysis}
\end{align}

{Based on~\eqref{eq_complexity_analysis}, we can see that the system parameters ($G$, $K$, $N$) affect the channel parameter estimation stage ($\mathcal{O}_2$, $\mathcal{O}_3$), the search grids ($Q_1$, $Q_2$, $Q_4$, $Q_5$, $Q_6$) affect the coarse estimations ($\mathcal{O}_2$, $\mathcal{O}_4$). The number of RISs $L$ increases the complexity for all the positioning steps, and the complexity increases quadratically in the positioning algorithm ($\mathcal{O}_4$, $\mathcal{O}_5$).
With more RISs introduced, the performance will improve based on the same system parameters (e.g., frequency and time resources, transmit power), owing to the extra geometrical information provided, as will be shown in Fig.~\ref{fig_multiple_RISs} and Fig.~\ref{fig_multiple_RIS_cdf}. However, the processing delay increases due to the high complexity of the positioning algorithms and possibly the increased $G$ due to the design of time-orthogonal beams. 
Considering the dimension of received signal symbols remains the same regardless of the number of RISs (only the channel changes), we anticipate machine learning and deep learning solutions have a huge potential to solve the scalability issue of a larger number of RISs and reduce the processing time of search-based algorithms.}

\subsection{{Extensions to More UEs}}
\label{sec_extension_more_UEs}
{Till now, we have discussed the multi-RIS-enabled sidelink positioning with a pair of transmitter and receiver UEs. For the scenarios with more than two UEs, two possible extensions can be considered; both require a coordinator to allocate resources and perform localization tasks~\cite{chen2023riss}:
\begin{enumerate}
    \item In the first scenario, one specific UE can be selected as the transmitter that broadcasts positioning reference signals to surrounding UEs. In this case, the structure and the algorithm mentioned in this work can be directly applied, and the positioning process can be performed at the receiver UEs in parallel without interference issues.
    \item In the second scenario, multiple UEs can play as transmitters and cooperate to obtain their positions (information exchange needed). To avoid interference between transmitting UEs, the time or frequency resources can be properly allocated for different UEs. Although with a higher overhead delay than the first scenario, the cooperation strategy can improve the system performance and can even work with a single RIS anchor with at least 3 UEs (see initial results reported in~\cite{ammous2023zero}).
\end{enumerate}

Due to the existence of blind areas, the information gains for different UEs from the same RIS will be different, which will be shown in the simulation section in terms of position error bound. In addition, the hardware capabilities and power constraints of different UEs require optimized power allocation and coordination. As a result, dedicated RIS profile optimization, resource allocation, and algorithm hyperparameter selection for multi-UE scenarios are needed in future work.
}

\section{Simulation Results}
\label{sec_simulation_results}
\subsection{Simulation Parameters}
We consider a 3D scenario with two single-antenna UEs and two RISs. The pilot signal $x_{g,k}$ has a constant amplitude. The unknown channel gains for the LOS path and the $\ell$-th RIS path are set as 
$|\rho_0| = \frac{\lambda_c}{4\pi d_0}$ 
and 
$|\rho_\ell| = \frac{\lambda_c^2}{16\pi^2 d_{\text{T},\ell}d_{\text{R},\ell}}$, both with random phase.
{When \acp{sp} are introduced, the channel gains are set as $|\rho_{0, m}| = \frac{\sqrt{4\pi c_{0,m}} \lambda_c}{16\pi^2 d_{\text{T},m}d_{\text{R},m}}$ and $|\rho_{\ell, m}| = \frac{\sqrt{4\pi c_{\ell, m}} \lambda_c^2}{64\pi^3 d_{\text{T},m}d_{\ell,m}d_{\text{R},\ell}}$, indicating respectively the propagation path of TX-SP-RX and TX-SP-RIS-RX, with $c_{\ell, m}$ as the \ac{rcs} coefficient of the $m$-th \ac{sp} in the $\ell$-th path.}
For channel parameter estimation, $N_{\text{FFT}} = 2^{10}$ is adopted, and the step size of the 2D grid search is $0.02$. In the positioning step, a step size $\unit[0.2]{m}$ is used to search around the true position (set as a $\unit[2\times2\times2]{m^2}$ area).
{The weight coefficients $w_{ij}$ in~\eqref{eq:candidate_UE} and $w_\ell$ in~\eqref{eq:coarse_position_costFun} are assigned with equal values, and $\Sigmam_{\eta_{\text{N}}}$ in~\eqref{eq:mle_position_estimation} is set as $\mathbf{I}$ by default.}
{The subcarrier spacing is chosen as $\unit[120]{kHz}$ ($\sim 3300$ subcarriers with a total $\unit[400]{MHz}$ bandwidth), and $K=512$ subcarriers are used for positioning pilot signals.}
The rest of the default simulation parameters can be found in Table~\ref{table_Simulation_parameters}. 



\vspace{-2mm}
\begin{table}[ht]
\centering
\caption{Default Simulation Parameters}
\renewcommand{\arraystretch}{0.85}
\begin{tabular} {c | c }
    \hthickline
    \textbf{Types} & \textbf{Simulation Parameters}\\
    \hline
    TX Position & {$\pv_\text{T} = [-2, -4, 0]^\top$}   \\
    \hline
    RX Position & {$\pv_\text{R} = [2, 3, 0]^\top$}   \\
    \hline
    RIS Positions & $\pv_1 = [-4, 0, 2]^\top$, $\pv_2=[4, 0, 2]^\top$ \\
    \hline
    RIS Orientation & $\ov_1 = [0, 0, 0]^\top$, $\ov_2=[\pi, 0, 0]^\top$ \\
    \hline
    RIS Size & {$N = 10\times 10$}   \\
    \hline
    Carrier Frequency & {$f_c = \unit[30]{GHz}$}
    \\
    \hline
    Bandwidth & {$W = \unit[400]{MHz}$}  \\
    \hline
    Number of Transmissions & {$G = 192$}  \\
    \hline
    Number of Subcarriers & {$K = 512$ (out of $3300$)} \\
    \hline
    Clock Offset & {$B = \unit[5]{m}$} \\
    \hline
    Average Transmission Power & {$P = 30$ dBm} \\
    \hline
    Noise PSD & {$N_0 = \unit[-173.855]{dBm/Hz}$} \\
    \hline
    Noise Figure & {$\unit[10]{dB}$} \\
    \hthickline
\end{tabular}
\renewcommand{\arraystretch}{1}
\label{table_Simulation_parameters}\vspace{-5mm}
\end{table}



\begin{figure}[h]
\centering
\begin{minipage}[b]{0.98\linewidth}
  \centering
%
%
\begin{tikzpicture}

\begin{axis}[%
width=2.5in,
height=1.6in,
at={(0in,0in)},
scale only axis,
xmin=12,
xmax=24,
xlabel style={font=\color{white!15!black},font=\footnotesize},
xlabel={Transmit Power [dBm]},
ymode=log,
ymin=0.001,
ymax=100,
yminorticks=false,
ylabel style={font=\color{white!15!black},font=\footnotesize},
ylabel={$\text{Error Bounds: d\ [m], }\xi\text{/}\zeta\text{ [no unit]}$},
axis background/.style={fill=white},
xmajorgrids,
ymajorgrids,
legend style={at={(1,1)},anchor=north east,legend cell align=left, align=left, font=\scriptsize, legend columns=3, draw=white!15!black}
]
\addplot [color=blue, line width=0.8pt, only marks, mark=o, mark options={solid, blue}]
  table[row sep=crcr]{%
10	86.5131187023842\\
12	53.602532744633\\
14	15.8409321972695\\
16	0.398581749048341\\
18	0.398979485566344\\
20	0.398979485566344\\
22	0.398979485566344\\
24	0.398979485566344\\
26	0.398979485566344\\
28	0.398979485566344\\
30	0.398979485566344\\
};
\addlegendentry{$d_1\ (\mathrm{coarse})$}

\addplot [color=blue, line width=0.8pt]
  table[row sep=crcr]{%
10	86.5010183562941\\
12	53.6149826985996\\
14	15.8392769979573\\
16	0.0296274067365484\\
18	0.0240233378225592\\
20	0.0182201504916143\\
22	0.0149734847618403\\
24	0.0123031637122592\\
26	0.00947542774011478\\
28	0.00769611804301984\\
30	0.00621691582290985\\
};
\addlegendentry{$d_1\ (\mathrm{fine})$}

\addplot [color=blue, dashed, line width=0.8pt]
  table[row sep=crcr]{%
10	0.0587058875537239\\
12	0.0466317440284716\\
14	0.0370409109162505\\
16	0.0294226413806846\\
18	0.0233712347888448\\
20	0.0185644316731498\\
22	0.0147462522395926\\
24	0.0117133645102746\\
26	0.00930425615412846\\
28	0.00739063336633139\\
30	0.00587058875537239\\
};
\addlegendentry{$\mathrm{DEB}_1$}

\addplot [color=red, only marks, line width=0.8pt, mark=square, mark options={solid, red}]
  table[row sep=crcr]{%
10	0.10265953886648\\
12	0.0583610808306282\\
14	0.029209357625802\\
16	0.0217857827214842\\
18	0.0204954172297588\\
20	0.0187422229687455\\
22	0.0179073582204801\\
24	0.0175102784445253\\
26	0.0176350989583207\\
28	0.0172324196150928\\
30	0.0170776365914464\\
};
\addlegendentry{$\xi_1\ (\mathrm{coarse})$}

\addplot [color=red, line width=0.8pt]
  table[row sep=crcr]{%
10	0.233448774424396\\
12	0.115306218967125\\
14	0.0671763653800493\\
16	0.0099601861922428\\
18	0.00727751066544348\\
20	0.00627557174382225\\
22	0.00506983477423951\\
24	0.00369584865017854\\
26	0.00309127529571345\\
28	0.00236332173714867\\
30	0.00201983027144503\\
};
\addlegendentry{$\xi_1\ (\mathrm{fine})$}

\addplot [color=red, dashed, line width=0.8pt]
  table[row sep=crcr]{%
10	0.0194939416256348\\
12	0.0154845882393087\\
14	0.0122998456415625\\
16	0.00977011467584346\\
18	0.00776067794351653\\
20	0.00616452561113715\\
22	0.00489665674660719\\
24	0.00388955270958323\\
26	0.0030895815376703\\
28	0.00245414184885438\\
30	0.00194939416256348\\
};
\addlegendentry{$\mathrm{EB}_{\xi_1}$}

\addplot [color=black, only marks, line width=0.8pt, mark=diamond, mark options={solid, black}]
  table[row sep=crcr]{%
10	0.240169895299025\\
12	0.183407331013547\\
14	0.0794757802354698\\
16	0.0165686162228615\\
18	0.0135916646703184\\
20	0.0121327424977926\\
22	0.00988326874057461\\
24	0.0087815330412924\\
26	0.00820390076138515\\
28	0.00751357908955296\\
30	0.00710492525308053\\
};
\addlegendentry{$\zeta_1\ (\mathrm{coarse})$}

\addplot [color=black, line width=0.8pt]
  table[row sep=crcr]{%
10	0.270336154903413\\
12	0.206445598924691\\
14	0.0764296108312448\\
16	0.0081598777983318\\
18	0.00660221730958667\\
20	0.00529570318766472\\
22	0.00426280674733098\\
24	0.00335925972814203\\
26	0.0026914961203446\\
28	0.00203512881205072\\
30	0.0017194333363447\\
};
\addlegendentry{$\zeta_1\ (\mathrm{fine})$}

\addplot [color=black, dashed, line width=0.8pt]
  table[row sep=crcr]{%
10	0.0168868822983385\\
12	0.0134137274060359\\
14	0.0106549024115093\\
16	0.00846348982369363\\
18	0.00672278893126148\\
20	0.00534010106419287\\
22	0.00424179305156957\\
24	0.00336937598671899\\
26	0.00267639047965288\\
28	0.00212593252513555\\
30	0.00168868822983385\\
};
\addlegendentry{$\mathrm{EB}_{\zeta_1}$}

\end{axis}

\end{tikzpicture}%
    \vspace{-.8 cm}
  \centerline{\small{(a) Channel parameter estimation}} \medskip
\end{minipage}
\vspace{-0.2cm}
\begin{minipage}[b]{0.98\linewidth}
\centering
%
%
\begin{tikzpicture}

\begin{axis}[%
width=2.5in,
height=1.6in,
at={(0in,0in)},
scale only axis,
xmin=12,
xmax=24,
xlabel style={font=\color{white!15!black},font=\footnotesize},
xlabel={Transmit Power [dBm]},
ymode=log,
ymin=0.03,
ymax=10,
yminorticks=false,
ylabel style={font=\color{white!15!black},font=\footnotesize},
ylabel={PEB [m] / CEB [m]},
axis background/.style={fill=white},
xmajorgrids,
ymajorgrids,
legend style={at={(1,1)},anchor=north east,legend cell align=left, align=left, font=\scriptsize, legend columns=3, draw=white!15!black}
]
\addplot [color=blue, only marks, line width=0.8pt, mark=o, mark options={solid, blue}]
  table[row sep=crcr]{%
10	1.50128113993892\\
12	1.08846939778221\\
14	0.804794018965635\\
16	0.795162869064078\\
18	0.7828193435273\\
20	0.768763772455952\\
22	0.775755106939632\\
24	0.77442804953844\\
26	0.770947298796026\\
28	0.752657991901676\\
30	0.76626933396205\\
};
\addlegendentry{$\mathbf{p}_\text{T}\ (\mathrm{coarse})$}

\addplot [color=blue, line width=0.8pt]
  table[row sep=crcr]{%
10	17.2982118741133\\
12	11.3297226171568\\
14	2.36340384261721\\
16	0.234908400256661\\
18	0.181861700709502\\
20	0.151342941071942\\
22	0.119399327644586\\
24	0.0947522786234257\\
26	0.0737876991758847\\
28	0.0584964020527782\\
30	0.0473765334079524\\
};
\addlegendentry{$\mathbf{p}_\text{T}\ (\mathrm{fine})$}

\addplot [color=blue, dashed, line width=0.8pt]
  table[row sep=crcr]{%
10	0.465527491529475\\
12	0.369781630600227\\
14	0.293727989865731\\
16	0.23331643568495\\
18	0.185329832477155\\
20	0.147212718671174\\
22	0.11693521895191\\
24	0.0928849460515579\\
26	0.0737811352271318\\
28	0.0586064389018554\\
30	0.0465527491561714\\
};
\addlegendentry{$\text{PEB}_\text{T}$}

\addplot [color=red, only marks, line width=0.8pt, mark=square, mark options={solid, red}]
  table[row sep=crcr]{%
10	5.29757826962371\\
12	4.70797818095349\\
14	1.38101053660444\\
16	1.07514457854924\\
18	1.06198837421432\\
20	1.04345527888914\\
22	1.06575404799872\\
24	1.06387915494621\\
26	1.05800104312918\\
28	1.03386897669405\\
30	1.05315842614606\\
};
\addlegendentry{$\mathbf{p}_\text{R}\ (\mathrm{coarse})$}

\addplot [color=red, line width=0.8pt]
  table[row sep=crcr]{%
10	18.4991394929788\\
12	11.1438527010225\\
14	2.23489195889961\\
16	0.215937194432035\\
18	0.163946313684149\\
20	0.13911847924964\\
22	0.104777056806008\\
24	0.0848156358566019\\
26	0.0645242245124102\\
28	0.0548038558193251\\
30	0.0417817813533451\\
};
\addlegendentry{$\mathbf{p}_\text{R}\ (\mathrm{fine})$}

\addplot [color=red, dashed, line width=0.8pt]
  table[row sep=crcr]{%
10	0.422387853239079\\
12	0.335514597841292\\
14	0.26650871822308\\
16	0.211695399693883\\
18	0.168155633130572\\
20	0.133570767224502\\
22	0.106099031738583\\
24	0.0842774565911734\\
26	0.0669439633199981\\
28	0.0531754802079575\\
30	0.0422387853232462\\
};
\addlegendentry{$\text{PEB}_\text{R}$}

\addplot [color=black, only marks, line width=0.8pt, mark=diamond, mark options={solid, black}]
  table[row sep=crcr]{%
10	3.3616185596302\\
12	3.60733128358475\\
14	0.737430302322289\\
16	0.617339086964357\\
18	0.606159557497349\\
20	0.575433753402918\\
22	0.592206220854673\\
24	0.594153310362328\\
26	0.58506210221709\\
28	0.568750933013981\\
30	0.577350850832943\\
};
\addlegendentry{$B\ (\mathrm{coarse})$}

\addplot [color=black, line width=0.8pt]
  table[row sep=crcr]{%
10	18.8392989508318\\
12	8.60558132139776\\
14	1.84365343750498\\
16	0.154932876953814\\
18	0.122076856705918\\
20	0.0941965617487693\\
22	0.078077751330569\\
24	0.0610579972711285\\
26	0.046901146276606\\
28	0.0371921410423762\\
30	0.0316951584665248\\
};
\addlegendentry{$B\ (\mathrm{fine})$}

\addplot [color=black, dashed, line width=0.8pt]
  table[row sep=crcr]{%
10	0.299045973754465\\
12	0.237540660473474\\
14	0.188685253504252\\
16	0.149878024339876\\
18	0.119052346480781\\
20	0.0945666402259775\\
22	0.0751169523873003\\
24	0.0596675161948077\\
26	0.0473955928084895\\
28	0.0376476575701293\\
30	0.0299045973777878\\
};
\addlegendentry{CEB}

\end{axis}
\end{tikzpicture}%
    \vspace{-0.8 cm}
  \centerline{\small{(b) Position estimation}} \medskip
\end{minipage}
\caption{{RMSE of the estimation results vs. derived CRBs: (a) channel parameter estimation, (b) position estimation, both are benchmarked by the derived CRBs. Coarse estimation results saturate at high transmission powers, whereas the refined results are able to attach the bounds.}}
\label{fig:estimator_vs_crb}
\vspace{-0.5cm}
\end{figure}


\subsection{Channel Parameters and Position Estimation Results}
\label{sec:sim_estimation_results}
\subsubsection{Positioning with a Limited Number of \acp{mpc}}
{We first evaluate the performance of the localization algorithm with a single \ac{sp} located at $[0, 2, 3]^\top\unit[]{m}$ (providing $M=3$ \acp{mpc} with $c_\text{RCS} = \unit[0.5]{m^2}$).} {Considering the multi-RIS-enabled 3D sidelink positioning discussed in this work is a novel scenario, no other benchmark algorithms are available. Hence, we use the derived CRB to evaluate the effectiveness of the proposed algorithms.} The channel parameter estimation and positioning results are shown in Fig.~\ref{fig:estimator_vs_crb} (a) and (b), respectively. It can be seen from both figures that the coarse estimations saturate to a certain level with the increased transmit power. However, when refinement processes are applied, the CRBs of channel parameters and state parameters can be attached when the transmit power is higher than $\unit[16]{dBm}$. 
{Note that when transmit power is low, coarse position results may perform better due to the constrained searching area, while the refinement process does not have such constraints (usually treated as no-information regions).} 
The results show the effectiveness of the derived bounds and the estimators at the high transmit power. Since refinement processes are involved, the tradeoff between positioning performance and complexity (e.g., the number of iterations) can be performed. In order to reduce the transmit power of the asymptotic region (i.e., $\unit[16]{dBm}$) for specific power-limited UE devices, RIS sizes can be increased, and antenna arrays at the UE side can be utilized for higher beamforming gain. However, the \ac{nf} channel model and orientation estimations for both UEs must be considered.

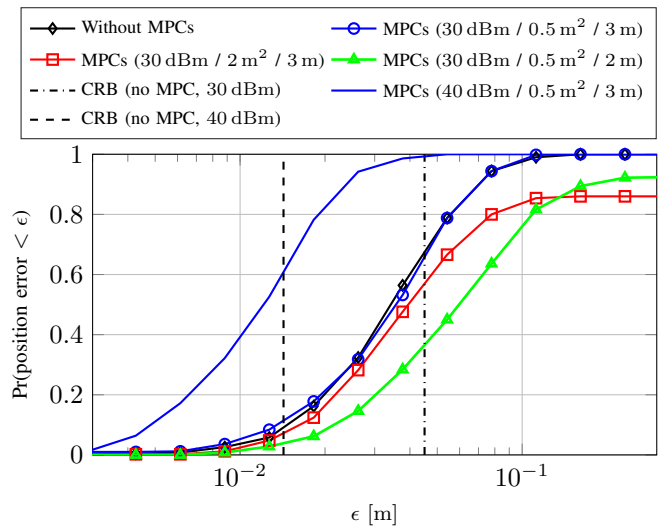
\begin{figure}[h]
\centering
\begin{minipage}[b]{0.98\linewidth}
%
%
\definecolor{mycolor1}{rgb}{0.00000,1.00000,1.00000}%
\definecolor{mycolor2}{rgb}{1.00000,0.00000,1.00000}%
\begin{tikzpicture}

\begin{axis}[%
width=7.5cm,
height=4cm,
at={(0in,0in)},
scale only axis,
xmode = log,
xmin=0.003,
xmax=0.3,
xlabel style={font=\color{white!15!black},font=\footnotesize},
xlabel={$\epsilon\  [\unit[]{m}]$},
ymin=0,
ymax=1,
yminorticks=true,
ylabel style={font=\color{white!15!black},font=\footnotesize},
ylabel={{Pr(position error $<\epsilon$)}},
axis background/.style={fill=white},
xmajorgrids,
ymajorgrids,
yminorgrids,
legend style={at={(1,1.05)}, anchor= south east, legend cell align=left, font=\scriptsize, align=left, legend columns=2, draw=white!15!black}
]
\addplot [color=black, mark=diamond, mark options={solid, black}, line width=0.8pt]
  table[row sep=crcr]{%
0.001	0\\
0.00143756376884548	0\\
0.00206658958949721	0\\
0.00297085431893443	0.002\\
0.00427079253141824	0.004\\
0.00613953660742272	0.00800000000000001\\
0.00882597538433136	0.026\\
0.0126879024372368	0.0580000000000001\\
0.0182396688464178	0.162\\
0.0262206870893498	0.324\\
0.0376939097538836	0.564\\
0.0541873989683141	0.788\\
0.0778978414848232	0.944\\
0.11198311458985	0.99\\
0.160982868256839	1\\
0.231423138810857	1\\
0.332685519626985	1\\
0.478256649435284	1\\
0.687524431437595	1\\
0.988360212830773	1\\
};
\addlegendentry{Without MPCs}

\addplot [color=blue, mark=o, mark options={solid, blue}, line width=0.8pt]
  table[row sep=crcr]{%
0.001	0.002\\
0.00143756376884548	0.002\\
0.00206658958949721	0.002\\
0.00297085431893443	0.01\\
0.00427079253141824	0.01\\
0.00613953660742272	0.012\\
0.00882597538433136	0.036\\
0.0126879024372368	0.084\\
0.0182396688464178	0.178\\
0.0262206870893498	0.318\\
0.0376939097538836	0.532\\
0.0541873989683141	0.788\\
0.0778978414848232	0.944\\
0.11198311458985	0.998\\
0.160982868256839	1\\
0.231423138810857	1\\
0.332685519626985	1\\
0.478256649435284	1\\
0.687524431437595	1\\
0.988360212830773	1\\
};
\addlegendentry{MPCs ($\unit[30]{dBm}$ / $\unit[0.5]{m^2}$ / $\unit[3]{m}$)}

\addplot [color=red, mark=square, mark options={solid, red}, line width=0.8pt]
  table[row sep=crcr]{%
0.001	0\\
0.00143756376884548	0\\
0.00206658958949721	0\\
0.00297085431893443	0\\
0.00427079253141824	0.002\\
0.00613953660742272	0.002\\
0.00882597538433136	0.012\\
0.0126879024372368	0.048\\
0.0182396688464178	0.124\\
0.0262206870893498	0.282\\
0.0376939097538836	0.476\\
0.0541873989683141	0.666\\
0.0778978414848232	0.8\\
0.11198311458985	0.854\\
0.160982868256839	0.86\\
0.231423138810857	0.86\\
0.332685519626985	0.86\\
0.478256649435284	0.86\\
0.687524431437595	0.86\\
0.988360212830773	0.86\\
};
\addlegendentry{MPCs ($\unit[30]{dBm}$ / $\unit[2]{m^2}$ / $\unit[3]{m}$)}

\addplot [color=green, mark=triangle, mark options={solid, green}, line width=1.pt]
  table[row sep=crcr]{%
0.001	0\\
0.00143756376884548	0\\
0.00206658958949721	0\\
0.00297085431893443	0\\
0.00427079253141824	0\\
0.00613953660742272	0\\
0.00882597538433136	0.00800000000000001\\
0.0126879024372368	0.028\\
0.0182396688464178	0.0620000000000001\\
0.0262206870893498	0.146\\
0.0376939097538836	0.284\\
0.0541873989683141	0.45\\
0.0778978414848232	0.636\\
0.11198311458985	0.816\\
0.160982868256839	0.894\\
0.231423138810857	0.922\\
0.332685519626985	0.924\\
0.478256649435284	0.924\\
0.687524431437595	0.948\\
0.988360212830773	0.996\\
};
\addlegendentry{MPCs ($\unit[30]{dBm}$ / $\unit[0.5]{m^2}$ / $\unit[2]{m}$)}

\addplot [color=black, dashdotted, line width=0.8pt]
  table[row sep=crcr]{%
0.0450807496484991	0\\
0.0450807496484991	1\\
};
\addlegendentry{CRB (no MPC, $\unit[30]{dBm}$)}

\addplot [color=blue, line width=0.8pt]
  table[row sep=crcr]{%
0.001	0\\
0.00143756376884548	0.002\\
0.00206658958949721	0.004\\
0.00297085431893443	0.016\\
0.00427079253141824	0.0639999999999999\\
0.00613953660742272	0.172\\
0.00882597538433136	0.322\\
0.0126879024372368	0.526\\
0.0182396688464178	0.782\\
0.0262206870893498	0.942\\
0.0376939097538836	0.986\\
0.0541873989683141	1\\
0.0778978414848232	1\\
0.11198311458985	1\\
0.160982868256839	1\\
0.231423138810857	1\\
0.332685519626985	1\\
0.478256649435284	1\\
0.687524431437595	1\\
0.988360212830773	1\\
};
\addlegendentry{MPCs ($\unit[40]{dBm}$ / $\unit[0.5]{m^2}$ / $\unit[3]{m}$)}

\addplot [color=black, dashed, line width=0.8pt]
  table[row sep=crcr]{%
0.0142557847517092	0\\
0.0142557847517092	1\\
};
\addlegendentry{CRB (no MPC, $\unit[40]{dBm}$)}

\end{axis}

\end{tikzpicture}%
    \vspace{-1.0 cm}
\end{minipage}
\caption{{Evaluation of the multipath on the estimator {in terms of CDF} (default multipath parameters are set as $P = \unit[30]{dBm}$, $c_\text{RCS}$ = $\unit[0.5]{m^2}$, $\text{height} = \unit[3]{m}$). We can see that different \acp{sp} properties (e.g., reflection coefficient, positions) affect positioning performance differently, and large transmit power could help to combat the multipath.}}
\label{fig:positioning_NLOS}
\vspace{-0.4cm}
\end{figure}

\subsubsection{The Effect of Multipath}
{We further explore the effect of multipath on sidelink positioning by creating two clusters of \acp{sp}. The first cluster has $5$ \acp{sp} that are distributed uniformly inside a disk on the x-y plane with a radius of $\unit[1]{m}$ centered at $[0, -3, 3]^\top \unit[]{m}$ (closer to the TX), providing $3\times5=15$ (one LOS and two RIS paths) \acp{mpc} for the LOS channel and TX-RIS channels. The second cluster is set similarly, centered at $[0, 2, 3]^\top \unit[]{m}$, to affect the LOS channel and RIS-RX channel at the RX side. A total number of $M = 30$ \acp{mpc} are generated in each realization to affect both the LOS channel and RIS channel. By default, the \ac{rcs} coefficient is set as $c_{\text{RCS}} = \unit[0.5]{m^2}$ for all the SPs, the phase of each \ac{mpc} path gain is chosen randomly for different realizations, and the transmit power is set as $P = \unit[30]{dBm}$ to make sure the estimator is working in the asymptotic region (i.e., close to the derived CRB). 
Since the position estimation performance of the TX and RX show a similar trend (which makes sense due to a large TX UE estimation error will affect the positioning of the RX UE), we focus on evaluating the estimation error of the RX. 

The \acp{cdf} of the estimation error less than a certain value and the CRB without considering multipath for 200 channel realizations are shown in Fig.~\ref{fig:positioning_NLOS}. 
When the \acp{mpc} with a small value of \ac{rcs} coefficients are introduced, the impact of multipath is limited (see the black curve with diamond markers and the blue curves with circular markers). However, the impact increases with an increased 
\ac{rcs} from $\unit[0.5]{m^2}$ to $\unit[2]{m^2}$ (see red curve with square markers). When changing the height of both cluster centers from $\unit[3]{m}$ to $\unit[2]{m}$, the \acp{mpc} are likely to make \ac{nlos} paths non-resolvable, resulting a worse performance under the same \ac{rcs} coefficient of $\unit[0.5]{m^2}$ (see the green curve with triangle markers). We can also see that a higher transmit power is needed to combat the effect of multipath to achieve satisfactory positioning accuracy. However, power consumption is critical on the UE side, and hence the development of the algorithm taking into account the multipath or the implementation of active \ac{ris} can be considered for future work.}
Considering the randomness of multipath, for simplicity, the simulation results in the following sections do not consider the effect of multipath, which lower bound the performance in the scenarios with multipath.

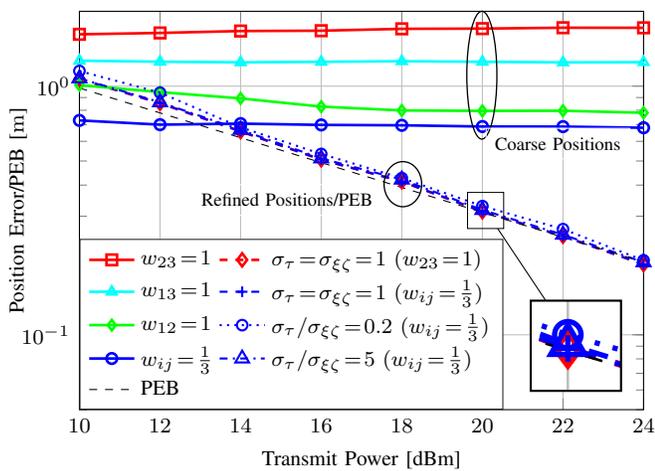
\begin{figure}[h]
\centering
%
\newcommand\ms{0.8}
\newcommand\lw{1.0}
\definecolor{mycolor1}{rgb}{0.00000,1.00000,1.00000}%
\begin{tikzpicture}[font=\footnotesize, spy using outlines={rectangle, magnification=2.5, size=0.3cm, connect spies}]
\begin{axis}[%
width=7.5cm,
height=5.3cm,
at={(0in,0in)},
scale only axis,
xmin=10,
xmax=24,
xlabel style={yshift=-0.0 ex},
xlabel={Transmit Power [dBm]},
ymode=log,
ymin=0.05,
ymax=2,
ylabel style={yshift=-2. ex},
ylabel={Position Error/PEB [$\unit[]{m}$]},
axis background/.style={fill=white},
xmajorgrids,
ymajorgrids,
legend columns=2, 
legend style={font=\footnotesize, at={(0, 0)}, anchor=south west, legend cell align=left, align=left, draw=white!15!black}
]
\addplot [color=red, line width=\lw pt, mark=square, mark options={solid, red}]
  table[row sep=crcr]{%
10	1.61626543392356\\
12	1.63684178026729\\
14	1.66602007132866\\
16	1.67106672286107\\
18	1.70029330247143\\
20	1.703794985478\\
22	1.71775590619561\\
24	1.71649757912669\\
26	1.72886423001034\\
28	1.72630797289615\\
30	1.73068103504817\\
};
\addlegendentry{${w_{23}}\!=\!1$}

\addplot [color=red, dashed, line width=\lw pt, mark size=2.5 pt, mark=diamond, mark options={solid, red}]
  table[row sep=crcr]{%
10	1.06733836482981\\
12	0.854367561842557\\
14	0.657804879167578\\
16	0.506523345960239\\
18	0.413502954852569\\
20	0.313057059134449\\
22	0.249530658966939\\
24	0.192001105759192\\
26	0.161391637901202\\
28	0.126183279871174\\
30	0.0961460232530284\\
};
\addlegendentry{$\sigma_{\tau} \!=\! \sigma_{\xi\zeta} \!=\! 1\ (w_{23}\!=\!1)$}

\addplot [color=mycolor1, line width=\lw pt, mark=triangle, mark options={solid, mycolor1}]
  table[row sep=crcr]{%
10	1.26537978474638\\
12	1.25406714152365\\
14	1.2465056063844\\
16	1.25279179377853\\
18	1.25970231852339\\
20	1.25287427162389\\
22	1.24602262198237\\
24	1.24873037053409\\
26	1.24352866191469\\
28	1.2323746940066\\
30	1.24281542147395\\
};
\addlegendentry{$w_{13}\!=\!1$}

\addplot [color=blue, dashed, mark=+, mark options={solid, blue}, line width=0.8 pt]
  table[row sep=crcr]{%
10	1.07042759189239\\
12	0.855142408822799\\
14	0.658161120572927\\
16	0.506621478127344\\
18	0.413499774335243\\
20	0.313163856495474\\
22	0.249476835681973\\
24	0.192009125113406\\
26	0.161390335974995\\
28	0.126162221012895\\
30	0.0961379153742007\\
};
\addlegendentry{$\sigma_{\tau} \!=\! \sigma_{\xi\zeta} \!=\! 1\ (w_{ij}\!=\!\frac{1}{3})$}

\addplot [color=green, line width=\lw pt, mark size=2 pt, mark=diamond, mark options={solid, green}]
  table[row sep=crcr]{%
10	1.01000232209897\\
12	0.94604930073816\\
14	0.89273166860556\\
16	0.827230772960148\\
18	0.798475850391974\\
20	0.795332943877414\\
22	0.796260263101614\\
24	0.781431724489027\\
26	0.780824468345871\\
28	0.794064705105964\\
30	0.761578664410381\\
};
\addlegendentry{$w_{12}\!=\!1$}

\addplot [color=blue, line width=0.8 pt, dotted, mark=o, mark options={solid, blue}]
  table[row sep=crcr]{%
10	1.15074826667723\\
12	0.94150324083346\\
14	0.683126462230254\\
16	0.533760378277095\\
18	0.426777022112351\\
20	0.329380787074528\\
22	0.266046820605742\\
24	0.199392616150283\\
26	0.167732680099291\\
28	0.138139336642425\\
30	0.102426635485195\\
};
\addlegendentry{$\sigma_{\tau}/\sigma_{\xi\zeta} \!=\! 0.2\ (w_{ij}\!=\!\frac{1}{3})$}

\addplot [color=blue, line width=1 pt, mark=o, mark size=2 pt, mark options={solid, blue}]
  table[row sep=crcr]{%
10	0.728193257113986\\
12	0.700064347423608\\
14	0.706841971525201\\
16	0.698392525481558\\
18	0.695940259616827\\
20	0.688030044781641\\
22	0.687469254319272\\
24	0.680289192280157\\
26	0.673908315436447\\
28	0.670021792708668\\
30	0.667827849266673\\
};
\addlegendentry{$w_{ij}\!=\!\frac{1}{3}$}

\addplot [color=blue, line width=0.8 pt, dashdotted, mark=triangle, mark size=3 pt, mark options={solid, blue}]
  table[row sep=crcr]{%
10	1.07462336026563\\
12	0.861554270132293\\
14	0.667169534230944\\
16	0.512222410300364\\
18	0.420946930913698\\
20	0.31769035374244\\
22	0.252683982519789\\
24	0.195837358349976\\
26	0.163763713711276\\
28	0.127526723874238\\
30	0.097373354916833\\
};
\addlegendentry{$\sigma_{\tau}/\sigma_{\xi\zeta} \!=\! 5\ ({w}_{ij}\!=\!\frac{1}{3})$}

\addplot [color=black, dashed]
  table[row sep=crcr]{%
10	0.982833715709978\\
12	0.780692570716027\\
14	0.620126151444504\\
16	0.492583711243175\\
18	0.391273149747553\\
20	0.310799310146107\\
22	0.246876667655267\\
24	0.196101107563245\\
26	0.155768646543933\\
28	0.123731433927773\\
30	0.0982833716333454\\
};
\addlegendentry{PEB}



\begin{scope}
    \spy[black, size=1.2cm] on (5.4, 2.65) in node [fill=none] at (6.6, 0.85);
\end{scope}
\end{axis}

\draw [black, line width= 0.5pt] 
(5.35, 4.45) 
ellipse [x radius = 0.2cm, y radius = 0.85cm];
\draw [black, line width=0.5pt] 
(4.3, 3.0) 
ellipse [x radius=0.25cm, y radius=0.3cm];
\node[below right, align=left, draw=white, font=\scriptsize]
at (5.4, 3.75) 
{Coarse Positions};
\node[below right, align=right, draw=white, font=\scriptsize]
at (1.5,3.0) 
{Refined Positions/PEB};

\end{tikzpicture}%
\vspace*{-1.cm}
\caption{{Evaluation of coarse and fine positioning results of the RX under different weight coefficients. We can see that the equal weight allocation to $w_{ij}$ achieves the best coarse estimation performance. However, even initialized using the worst coarse estimation (i.e., $w_{23}=1$), the refined results are not largely affected. In the refined results, the positions initialized using $w_{ij}=\frac{1}{3}$, $\sigma_\tau/\sigma_{\xi\zeta}=1$ is close to the PEB and perform better than the coefficient selection relying more on delay (i.e., $\sigma_\tau/\sigma_{\xi\zeta}=0.2$ where the delay is assumed to have less error).}}
\label{fig_weight_selection}
\vspace{-0.2cm}
\end{figure}

\begin{figure*}[h]
\centering
\begin{minipage}[h]{0.95\linewidth}
\centering
\centerline{\includegraphics[width=1.\linewidth]{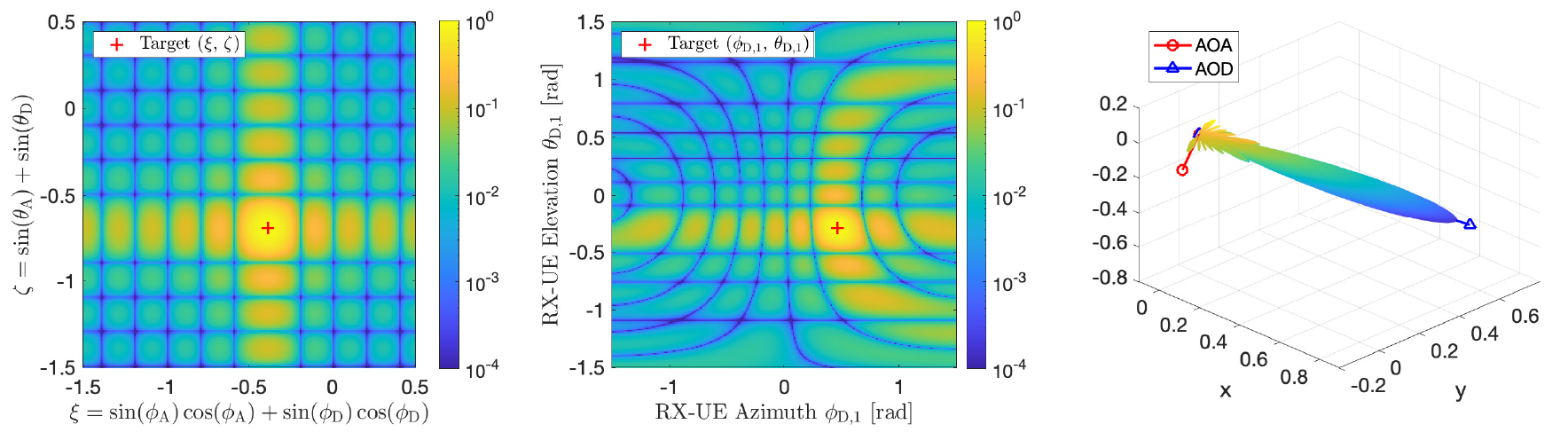}}
\small
\centerline{\hspace{0.4cm} (a)-1 \hspace{5.2cm} (a)-2 \hspace{5.4cm} (a)-3}
\vspace{0.2cm}
\centerline{(a) DIR beam $\omegav_1$} \medskip
\normalsize
\end{minipage}

\begin{minipage}[h]{0.95\linewidth}
\centering
\centerline{\includegraphics[width=1.\linewidth]{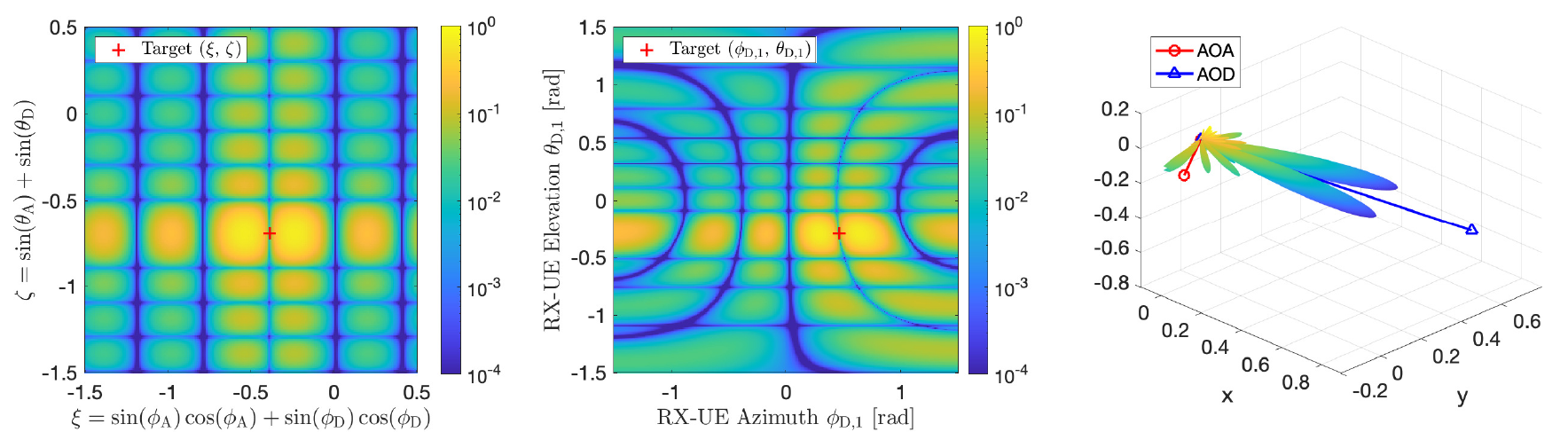}}
\small
\centerline{\hspace{0.4cm} (b)-1 \hspace{5.2cm} (b)-2 \hspace{5.4cm} (b)-3}
\vspace{0.2cm}
\centerline{(b) DER beam $\omegav_2$} \medskip
\normalsize
\vspace{-0.2cm}
\end{minipage}
\caption{The beam pattern of (a) DIR beam $\omegav_1$ and (b) DER beam $\omegav_2$. The figures in column-1 show the {normalized} radiation patterns of two beams by changing $\xi$ and $\zeta$, and the DIR beam in (a)-1 reaches the maximum at ($\xi = -0.4443$, $\zeta = -0.5039$). The figures in column-2 and column-3 visualize the radiation patterns (in 2D and 3D, respectively) by changing the azimuth and elevation of the receiver $\varphiv_{\text{A},1}$, with a fixed $\varphiv_{\text{A},1} = [-1.1071, -0.2200]^\top$ (rad), the DIR beam in (a)-2 reaches the maximum at $\varphiv_{\text{D},1} = [0.4636, -0.2898]^\top$ (rad).}
\label{fig_RIS_profile_visualization}
\end{figure*}

\subsubsection{{Evaluation of Weight Coefficients Selection}}
\label{sec_weighting_coefficients_3RIS}

{In the scenarios with 2 RISs, there is only one position candidate $\check{\pv}_\text{R} = \check{\pv}_{12}$ ($w_{12}=1$ in~\eqref{eq:candidate_UE}). When extending to 3 RISs, the choice of $w_{12}$, $w_{13}$ and $w_{23}$ affects the positioning performance. We visualize four weight selection results, namely, $w_{12}=1$ (i.e., use $\check{\pv}_{12}$ directly), $w_{13}=1$, $w_{23}=1$, and $w_{12} = w_{23}=w_{13}=\frac{1}{3}$ (denoted as $w_{ij}=\frac{1}{3}$), as shown in Fig.~\ref{fig_weight_selection}. We notice that $w_{ij}=\frac{1}{3}$ performs better in the coarse position estimate than other coefficients mentioned before (see solid curves). However, the refined position RMSE with $w_{23}=1$ and $w_{ij}=\frac{1}{3}$ coefficient allocation perform almost the same, with the default $\Sigmam_{\eta_{\text{N}}}=\mathbf{I}$ (denoted by $\sigma_{\tau}=\sigma_{\xi\zeta}=1$). It is also shown that relying more on delay estimation (i.e., assigning $\sigma_{\tau}^2$ to the delay-related entries and $\sigma_{\tau}^2$ to the rest of the entries in the diagonal $\Sigmam_{\eta_{\text{N}}}$) yield worse performance. Note that the fixed weights used in Fig.~\ref{fig_weight_selection} are not optimal, and the weight optimization problem is left for future work.
}

\subsection{Evaluation of RIS Profiles}
\subsubsection{Visualization of DIR and DER beams}
Based on the simulation parameters in Table~\ref{table_Simulation_parameters}, we first visualize the radiation patterns (i.e., the equivalent RIS gain $|\omegav^\top \av_\text{R}(\xi, \zeta) |$) of DIR beam $\omegav_1$ and DER beam $\omegav_2$ obtained from~\eqref{eq:sum_diff_beam_1} and~\eqref{eq:sum_diff_beam_2} for the first RIS $\pv_{1}$.
By changing the spatial frequencies $\xi$ and $\zeta$, the radiation patterns of two beams are shown in Fig.~\ref{fig_RIS_profile_visualization} (a)-1 and (b)-1. If we assume the position of the TX is known and fix the transmitter angles as $\varphiv_{\text{A},1}$, the 2D radiation patterns of the two beams are visualized in Fig.~\ref{fig_RIS_profile_visualization} (a)-2, (b)-2, and the 3D radiation patterns are visualized in Fig.~\ref{fig_RIS_profile_visualization} (a)-3, (b)-3, respectively. We can see from the figures that the DIR beam maximizes the \ac{snr} of the TX-RX link, while the DER beams are split at the dimension of $\xi$ and $\zeta$ compared with the DIR beam.
The DER beam $\omegav_3$ (derivation with respect to $\zeta$) shows a similar pattern to Fig.~\ref{fig_RIS_profile_visualization} (b)-1, by splitting the beam from the $\zeta$ axis, which is not plotted.

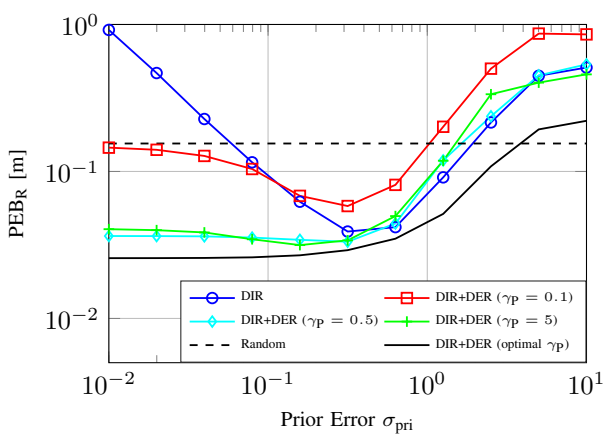
\begin{figure}[t]
\begin{minipage}[t]{0.98\linewidth}
\centering
%
%
\definecolor{mycolor1}{rgb}{0.00000,1.00000,1.00000}%
\definecolor{mycolor2}{rgb}{1.00000,0.00000,1.00000}%
\begin{tikzpicture}

\begin{axis}[%
width=2.5in,
height=4.5cm,
at={(0in,0in)},
scale only axis,
xmin=0.01,
xmax=10,
xmode = log,
xlabel style={font=\color{white!15!black},font=\footnotesize},
xlabel={Prior Error $\sigma_{\text{pri}}$},
ymin=0.005,
ymax=1,
ymode=log,
yminorticks=false,
ylabel style={font=\color{white!15!black},font=\footnotesize},
ylabel={$\text{PEB}_\text{R}$ [m]},
axis background/.style={fill=white},
xmajorgrids,
ymajorgrids,
legend style={at={(1,0)}, anchor=south east,legend cell align=left, font=\tiny, align=left, draw=white!15!black, legend columns=2}
]
\addplot [color=blue, mark=o, line width=0.7pt, mark options={solid, blue}]
  table[row sep=crcr]{%
0.01	0.918218228628026\\
0.0199526231496888	0.467402404915849\\
0.0398107170553497	0.227453668400114\\
0.0794328234724281	0.115425069449969\\
0.158489319246111	0.0622535374397418\\
0.316227766016838	0.0390218199863891\\
0.630957344480193	0.0417965419500264\\
1.25892541179417	0.0911587546544416\\
2.51188643150958	0.215404870756842\\
5.01187233627272	0.447478785054419\\
10	0.509281743063723\\
};
\addlegendentry{DIR}

\addplot [color=red, mark=square, line width=0.7pt, mark options={solid, red}]
  table[row sep=crcr]{%
0.01	0.145222919791303\\
0.0199526231496888	0.140428336803404\\
0.0398107170553497	0.127541110172145\\
0.0794328234724281	0.103938460454019\\
0.158489319246111	0.0682460715814379\\
0.316227766016838	0.0581769537548144\\
0.630957344480193	0.0810617417122793\\
1.25892541179417	0.201395163602589\\
2.51188643150958	0.500624592414197\\
5.01187233627272	0.866714209844032\\
10	0.856187319624162\\
};
\addlegendentry{DIR+DER ($\gamma_\text{P} = 0.1$)}

\addplot [color=mycolor1, mark=diamond, line width=0.7pt, mark options={solid, mycolor1}]
  table[row sep=crcr]{%
0.01	0.0363742056501761\\
0.0199526231496888	0.0363362565486738\\
0.0398107170553497	0.0361846934236847\\
0.0794328234724281	0.0355016737198167\\
0.158489319246111	0.034274102284504\\
0.316227766016838	0.0332540419489672\\
0.630957344480193	0.0442547520011064\\
1.25892541179417	0.118958961059885\\
2.51188643150958	0.237327846801394\\
5.01187233627272	0.450552758206164\\
10	0.535224550766476\\
};
\addlegendentry{DIR+DER ($\gamma_\text{P} = 0.5$)}

\addplot [color=green, mark=+, line width=0.7pt, mark options={solid, green}]
  table[row sep=crcr]{%
0.01	0.0404867849111126\\
0.0199526231496888	0.0398783790983836\\
0.0398107170553497	0.038475472699799\\
0.0794328234724281	0.034577469813188\\
0.158489319246111	0.0315558766782854\\
0.316227766016838	0.0339925789642226\\
0.630957344480193	0.049709997155062\\
1.25892541179417	0.118273495001455\\
2.51188643150958	0.335272368798396\\
5.01187233627272	0.402156708110215\\
10	0.457651416350936\\
};
\addlegendentry{DIR+DER ($\gamma_\text{P} = 5$)}

\addplot [color=black, dashed, line width=0.7pt]
  table[row sep=crcr]{%
0.01	0.154968485387534\\
0.0199526231496888	0.154968485387534\\
0.0398107170553497	0.154968485387534\\
0.0794328234724281	0.154968485387534\\
0.158489319246111	0.154968485387534\\
0.316227766016838	0.154968485387534\\
0.630957344480193	0.154968485387534\\
1.25892541179417	0.154968485387534\\
2.51188643150958	0.154968485387534\\
5.01187233627272	0.154968485387534\\
10	0.154968485387534\\
};
\addlegendentry{Random}

\addplot [color=black, line width=0.7pt]
  table[row sep=crcr]{%
0.01	0.0257166751715556\\
0.0199526231496888	0.0257342942724258\\
0.0398107170553497	0.0258036255010641\\
0.0794328234724281	0.0260440227932588\\
0.158489319246111	0.0268911770979311\\
0.316227766016838	0.0291411108941801\\
0.630957344480193	0.0348845755170169\\
1.25892541179417	0.0514381746904365\\
2.51188643150958	0.108314575955712\\
5.01187233627272	0.193304358157782\\
10	0.221129237001809\\
};
\addlegendentry{DIR+DER (optimal $\gamma_\text{P}$)}


\end{axis}

\end{tikzpicture}%
\vspace{-0.8 cm}
\caption{The evaluation of different prior error levels on positioning with different RIS profile designs. We can see that by using a DIR codebook, the positioning error bounds decrease and then increase when reducing the prior error level, and this effect can be mitigated by power control.}
\label{fig_RIS_profiles_prior}
\end{minipage}
\end{figure}
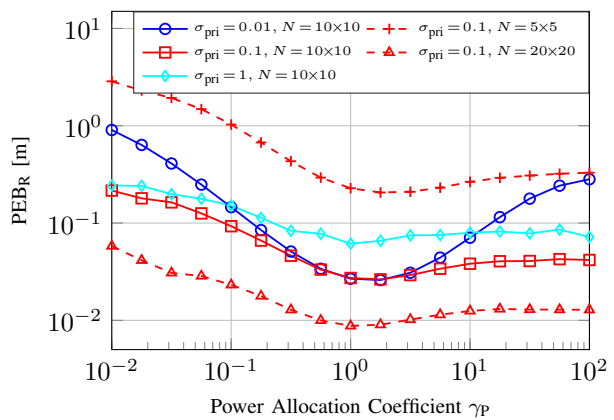
\begin{figure}[t]
\begin{minipage}[t]{0.98\linewidth}
\centering
%
%
\definecolor{mycolor1}{rgb}{0.00000,1.00000,1.00000}%
\definecolor{mycolor2}{rgb}{1.00000,0.00000,1.00000}%
\begin{tikzpicture}

\begin{axis}[%
width=2.5in,
height=4.5cm,
at={(0in,0in)},
scale only axis,
xmin=0.01,
xmax=100,
xmode = log,
xlabel style={font=\color{white!15!black},font=\footnotesize},
xlabel={Power Allocation Coefficient $\gamma_{\text{P}}$},
ymin=0.005,
ymax=15,
ymode=log,
yminorticks=false,
ylabel style={font=\color{white!15!black},font=\footnotesize},
ylabel={$\text{PEB}_\text{R}$ [m]},
axis background/.style={fill=white},
xmajorgrids,
ymajorgrids,
legend style={at={(1,1)}, anchor=north east,legend cell align=left, font=\tiny, align=left, draw=white!15!black, legend columns=2}
]
\addplot [color=blue, mark=o, line width=0.7pt, mark options={solid, blue}]
  table[row sep=crcr]{%
0.01	0.904521534330284\\
0.0177827941003892	0.63478538631615\\
0.0316227766016838	0.409503757579986\\
0.0562341325190349	0.248326115222642\\
0.1	0.145657852362741\\
0.177827941003892	0.0846340303107656\\
0.316227766016838	0.0509661606211919\\
0.562341325190349	0.0338662576962139\\
1	0.0267589161218137\\
1.77827941003892	0.0259298014336519\\
3.16227766016838	0.0308235934196885\\
5.62341325190349	0.0440909274433338\\
10	0.0706035607803841\\
17.7827941003892	0.114968147487771\\
31.6227766016838	0.178117210547176\\
56.2341325190349	0.242011653558318\\
100	0.28195152653735\\
};
\addlegendentry{$\sigma_{\text{pri}}\!=\!0.01, N\!=\!10\!\!\times\!\! 10$}

\addplot [color=red, dashed, line width=0.7pt, mark=+, mark options={solid, red}]
  table[row sep=crcr]{%
0.01	2.86124916631659\\
0.0177827941003892	2.30297614214841\\
0.0316227766016838	1.93290690102061\\
0.0562341325190349	1.48241009624327\\
0.1	1.02934488763526\\
0.177827941003892	0.673070950635365\\
0.316227766016838	0.431724855991769\\
0.562341325190349	0.292906300029651\\
1	0.227778616492719\\
1.77827941003892	0.206151420188041\\
3.16227766016838	0.209635714178467\\
5.62341325190349	0.231039196668893\\
10	0.26530334578633\\
17.7827941003892	0.291702119257105\\
31.6227766016838	0.307460544775436\\
56.2341325190349	0.321177184644911\\
100	0.329164498970705\\
};
\addlegendentry{$\sigma_{\text{pri}}\!=\!0.1, N\!=\!5\!\!\times\!\! 5$}

\addplot [color=red, mark=square, line width=0.7pt, mark options={solid, red}]
  table[row sep=crcr]{%
0.01	0.215842408002819\\
0.0177827941003892	0.179494754975106\\
0.0316227766016838	0.163176904076245\\
0.0562341325190349	0.125635673090007\\
0.1	0.0927788080249611\\
0.177827941003892	0.0660325526430829\\
0.316227766016838	0.0461289714120673\\
0.562341325190349	0.0331988314006826\\
1	0.0272260965884278\\
1.77827941003892	0.0263470911012587\\
3.16227766016838	0.0291781178798407\\
5.62341325190349	0.033886421239511\\
10	0.0381619782769032\\
17.7827941003892	0.0404596852571792\\
31.6227766016838	0.0406905531290478\\
56.2341325190349	0.0425415257061215\\
100	0.0417236633507587\\
};
\addlegendentry{$\sigma_{\text{pri}}\!=\!0.1, N\!=\!10\!\!\times\!\! 10$}

\addplot [color=red, dashed, line width=0.7pt, mark=triangle, mark options={solid, red}]
  table[row sep=crcr]{%
0.01	0.0579162230449081\\
0.0177827941003892	0.0414047579975162\\
0.0316227766016838	0.0307199563846571\\
0.0562341325190349	0.0285299641831158\\
0.1	0.0231519547940919\\
0.177827941003892	0.0177673639846759\\
0.316227766016838	0.0128289602095385\\
0.562341325190349	0.00997016744107829\\
1	0.00875388815373437\\
1.77827941003892	0.00901132355449595\\
3.16227766016838	0.0101761497476183\\
5.62341325190349	0.0114207949966697\\
10	0.0124797744383646\\
17.7827941003892	0.0130746641533555\\
31.6227766016838	0.012908150683727\\
56.2341325190349	0.012873335887883\\
100	0.012830724885758\\
};
\addlegendentry{$\sigma_{\text{pri}}\!=\!0.1, N\!=\!20\!\!\times\!\! 20$}

\addplot [color=mycolor1, mark=diamond, line width=0.7pt, mark options={solid, mycolor1}]
  table[row sep=crcr]{%
0.01	0.242825439231786\\
0.0177827941003892	0.241020616181188\\
0.0316227766016838	0.198004619399491\\
0.0562341325190349	0.177764135972638\\
0.1	0.151300230543826\\
0.177827941003892	0.113000178659281\\
0.316227766016838	0.0832102636937878\\
0.562341325190349	0.0776251860518739\\
1	0.0613453264480001\\
1.77827941003892	0.0657214114810711\\
3.16227766016838	0.0745688751393564\\
5.62341325190349	0.0753470172291155\\
10	0.0794818181968201\\
17.7827941003892	0.081797682633952\\
31.6227766016838	0.0783856884254696\\
56.2341325190349	0.085587804219379\\
100	0.0718562477438121\\
};
\addlegendentry{$\sigma_{\text{pri}}\!=\!1, N\!=\!10\!\!\times\!\! 10$}

\end{axis}

\end{tikzpicture}%
\vspace{-0.8 cm}
\caption{The evaluation of different power allocation coefficients for different prior error levels and RIS sizes. We notice that power control is important when the prior error level is small, and the selection of the optimal coefficient also depends on the RIS size.}
\label{fig_RIS_profiles_power_allocation}
\end{minipage}
\end{figure}

\subsubsection{The Effect of Prior Error Level on RIS Profile Design}
To evaluate the effect of prior error level on the $\text{PEB}_\text{R}$ for different RIS profile designs, we assume the covariance matrices of the prior information are set as $\bar\Sigmam_{\sv_\text{N}} = \sigma^2_\text{pri}\mathbf{I}\in \mathbb{R}^{3\times 3}$, for simplicity. Benchmarked by the random RIS profile (black dashed curve), the PEBs for the DIR codebook, and DIR+DER codebooks with different power allocations are shown in Fig.~\ref{fig_RIS_profiles_prior}. We can see from the figure that both DIR and DIR+DER RIS profiles do not help when the prior error level is high. With more accurate prior information, the DIR profile can largely reduce the PEB. However, when the prior error is too small, the RIS profiles based on the DIR beams are configured to beamforming to a small area and provide less spatial diversity. In an extreme case, the positioning task cannot be completed with all the beams pointing to a single point. When adopting the DIR+DER profiles, however, this phenomenon can be mitigated by choosing a proper power allocation coefficient $\gamma_\text{P}$.

We further evaluate the effect of power allocation coefficients on the positioning error bound, as shown in Fig.~\ref{fig_RIS_profiles_power_allocation}. For a fixed RIS size (i.e., $10 \times 10$), {the performance improved by power allocation is limited when the prior error level is high} (cyan curve with diamond markers), and becomes crucial with a small error level (blue curve with circle markers). We can also see that the optimal coefficient slightly shifts from left (red triangle) to right (red cross) with the increase of RIS sizes, which is due to the narrow beamwidth requiring accurate prior information.


\begin{figure*}[t]
\begin{minipage}[b]{0.99\linewidth}
\begin{center}
\end{center}
\centering
    \centerline{\includegraphics[width=1.\linewidth]{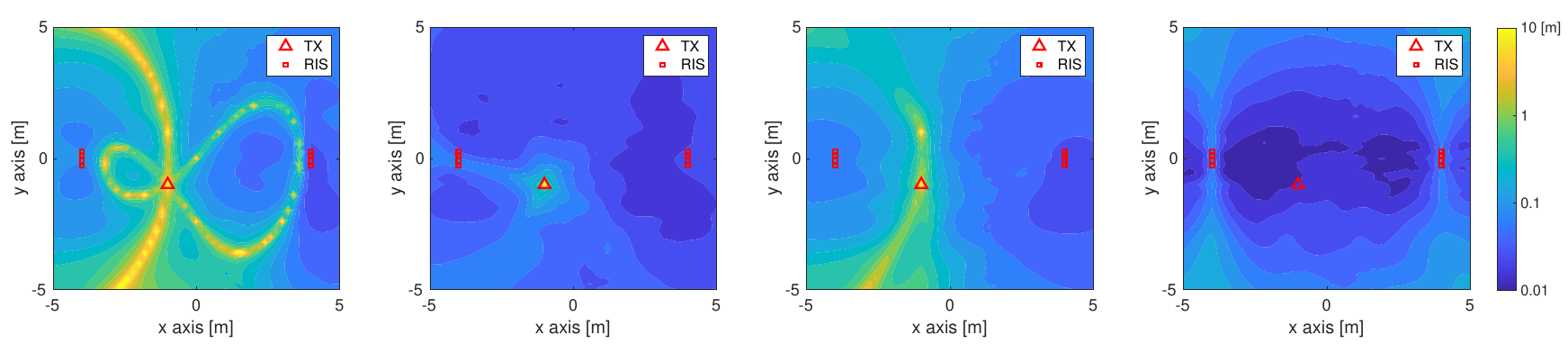}}
    \small
    \centerline{(a)-1 \hspace{3.4cm} (a)-2 \hspace{3.4cm} (a)-3 \hspace{3.4cm} (a)-4 \hspace{0.2cm}}
    \vspace{0.15cm}
    \centerline{(a) Scenario 1: Two RISs are parallel facing each other.}
    \normalsize
\end{minipage}
\begin{minipage}[b]{0.99\linewidth}
    \centering
    \centerline{\includegraphics[width=1.\linewidth]{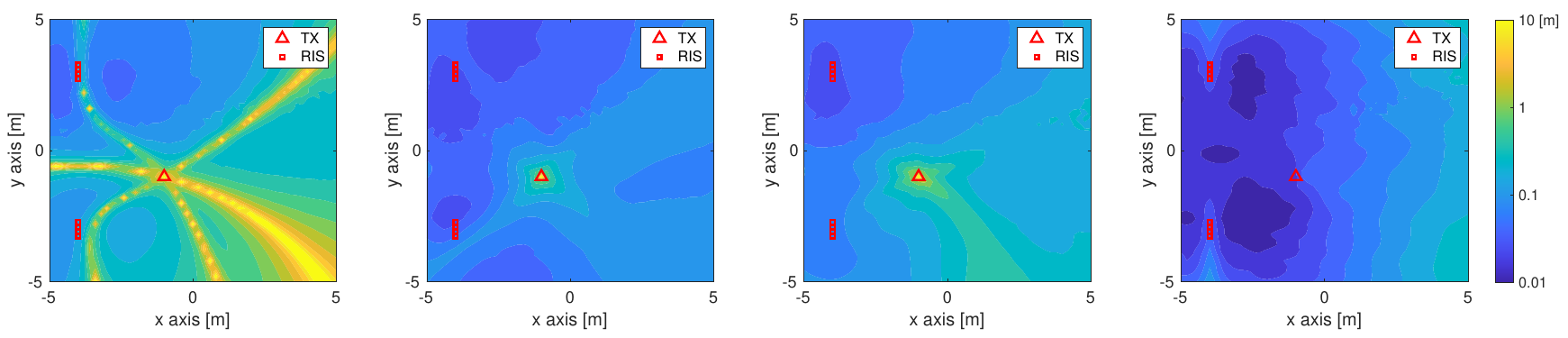}}
    \small
    \centerline{(b)-1 \hspace{3.4cm} (b)-2 \hspace{3.4cm} (b)-3 \hspace{3.4cm} (b)-4 \hspace{0.2cm}}
    \vspace{0.15cm}
    \centerline{(b) Scenario 2: Two RISs are located on the same plane facing the +x axis.}
    \normalsize
\end{minipage}
\begin{minipage}[b]{0.99\linewidth}
    \centering
    \centerline{\includegraphics[width=1.\linewidth]{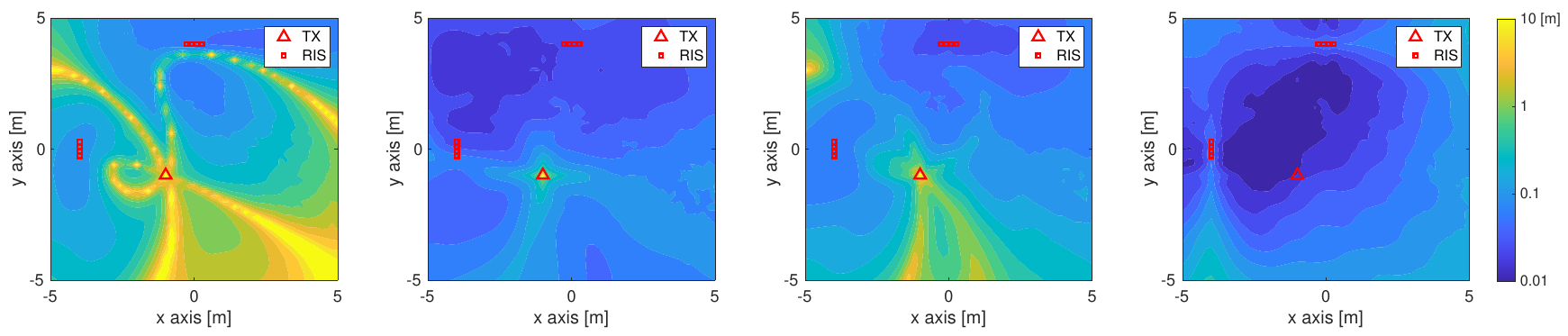}}
    \small
    \centerline{(c)-1 \hspace{3.4cm} (c)-2 \hspace{3.4cm} (c)-3 \hspace{3.4cm} (c)-4 \hspace{0.2cm}}
    \vspace{0.15cm}
    \centerline{(c) Scenario 3: Two RISs are perpendicular to each other.}
\end{minipage}
\caption{Visualization of PEBs of RX at different positions while the TX is fixed (random RIS codebooks). Three scenarios are considered where two RISs (same height) are located $\unit[2]{m}$ above the UEs (same height): a) two RIS are parallel facing each other. b). Two RISs are located parallel to the y-axis. c). Two RISs are forming an L shape facing the positive of the X axis and the negative of the Y axis (e.g., a corner). For each benchmark scenario (column-1), PEBs with known clock offset (column-2), PEBs with known z-axis (column-3), and {PEBs with known TX position (column-4) are also visualized}.}\vspace{-3mm}
\label{fig_heatmap}
\end{figure*}

\subsection{Localizability Discussion}
\subsubsection{PEB visualization of different RIS layouts}
Based on the analysis in Section~\ref{sec:problem_statement}, at least two RISs are needed to enable sidelink positioning, under the far-field assumption. However, this may not always work, and the localizability also depends on the state of the RIS anchors and the UEs. We visualize the PEBs of the RX UE (with TX UE position fixed) on a 2D x-y plane where both UEs have an unknown but fixed height, RISs are $\unit[2]{m}$ above both UEs, and the TX UE is located at $[-1, -1, 0]^\top$. Without loss of generality, random RIS codebooks and $P = \unit[30]{dBm}$ are adopted. Three different RIS layouts are considered: (a) two RISs at different locations ($[-4, 0, 2]^\top$ and $[4, 0, 2]^\top$) facing each other, (b) two RISs are located on the same y-z plane ($[-4, -3, 2]^\top$ and $[-4, 3, 2]^\top$) facing the same direction, and (c) two RISs are perpendicular to each other ($[-4, 0, 2]^\top$ and $[0, 4, 2]^\top$). The results are shown in Fig.~\ref{fig_heatmap} with different assumptions. Benchmarked by the default setup (column-1), the PEB for perfect synchronization (column-2), UEs with known heights (column-3), and positioning of RX with known TX position (column-4) are visualized. The positions behind the RIS are also plotted, as there exist certain types of RIS that are able to refract the signals rather than reflect them.

We can see from Fig.~\ref{fig_heatmap} that the blind areas exist (yellow area), where the positioning cannot be done or will yield poor performance. However, with extra information, such as clock offset (column-2) or known UE heights (column-3), the blind area can be largely reduced. We also notice that the blind areas in the 3D space are changing continuously (see column-1 and column-4). Since the derivation of PEBs involves a high dimension of parameters such as the RIS positions/orientations and UE positions, it would be challenging to derive a closed-form solution to avoid these areas.

\begin{figure*}[t]
\centering
\centering
\centerline{\includegraphics[width=1.\linewidth]{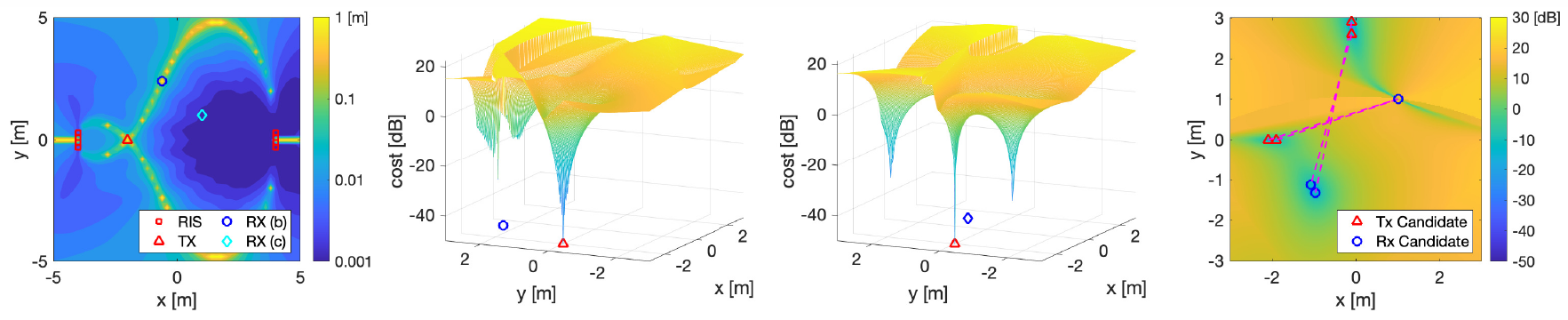}}
\small
\centerline{ \hspace{0.1cm} (a) \hspace{4.3cm} (b) \hspace{4.1cm} (c) \hspace{3.8cm} (d) \hspace{0.2cm}}
\normalsize
\vspace{-0.1cm}
\caption{Interpretation of the blind area. (a) PEB Heatmap of a 2D scenario where RISs and UEs are on the same X-Y plane and the UEs have known heights; (b) Cost function for the RX located at $[-0.61, 2.41, 0]^\top$ (blind area); (c) Cost function for the RX located at $[1.01, 1.01, 0]^\top$ (non-blind area); (d) Candidate TX/RX pairs at the locations around the local minima in (c).}
\label{fig_blind_area_interpretation}
\vspace{-0.5cm}
\end{figure*}

\begin{figure*}[h]
\centering
\centerline{\includegraphics[width=1.\linewidth]{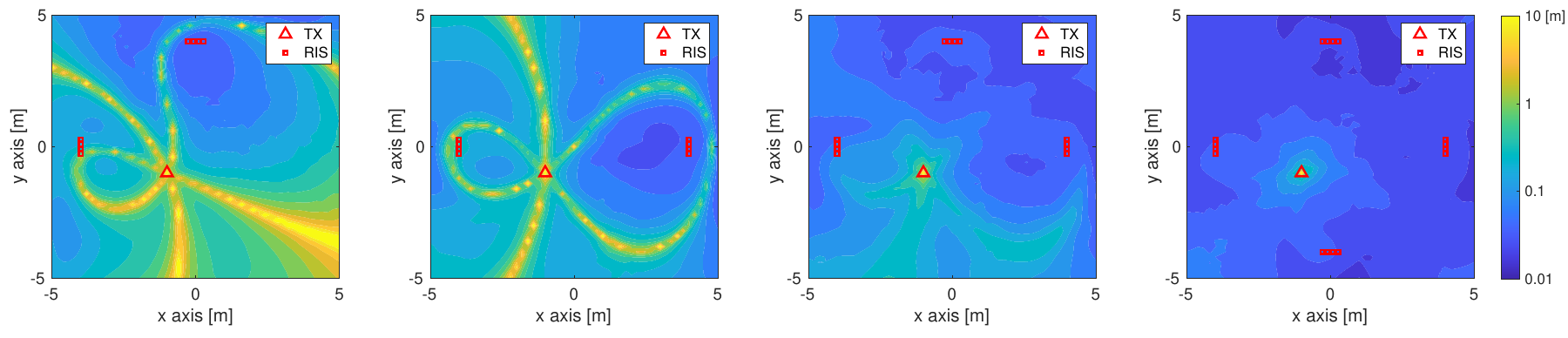}}
\small
\centerline{(a) \hspace{3.8cm} (b) \hspace{3.8cm} (c) \hspace{3.8cm} (d) \hspace{0.2cm}}
\normalsize
\vspace{-0.3 cm}
\caption{Visualization of RX PEBs for different numbers of RISs (with $30^\circ$ tilting down). Subfigures (a) and (b) show that the RIS orientations affect the PEB (compared with the benchmarks in Fig.~\ref{fig_heatmap} (a)-1, (c)-1). With more RISs, the areas with poor positioning performance can be largely reduced.}
\label{fig_multiple_RISs}
\vspace{-0.5cm}
\end{figure*}

\subsubsection{Interpretation of the Blind Area}
For a fixed TX UE position, worse performance (yellow area) happens in the location where the surrounding RX positions can provide similar geometrical information. To visualize this, we plot the noise-free cost function, defined in~\eqref{eq:coarse_position_costFun}, for a 2D scenario. In the localizable location, the cost function shows a clear, sharp global optimal. In the blind area, the cost of the optimal point is similar to the surrounding positions, and the same level of noise will cause a larger estimation error compared with the first scenario. We further choose several surrounding candidate RX UE positions and find the corresponding optimal TX that minimizes the cost function, as shown in Fig.~\ref{fig_blind_area_interpretation} (d). All these pairs provide similar spatial frequencies and TDOA observations as the ground truth TX/RX pair, and erroneous estimations are likely to happen at these positions.
In order to avoid the effect of local optima, global optimization methods can be adopted. In addition, prior location information can also effectively eliminate the local minima.
As mentioned earlier, for a fixed setup, the blind areas can be reduced by adopting a round-trip estimation to remove the clock offset, using geometric constraints (e.g., known UE height). In the next, we discuss the scenarios with more than two RISs.

\subsubsection{Evaluation of More RISs and TX/RX Pairs}

\begin{figure}[h]
\centering
  \centering
%
%
\begin{tikzpicture}

\begin{axis}[%
width=2.99in,
height=1.8in,
at={(0in,0in)},
scale only axis,
xmode=log,
xmin=0.01,
xmax=1,
xminorticks=true,
xlabel style={font=\color{white!15!black},font=\footnotesize},
xlabel={$\epsilon$ [m]},
ymin=0.,
ymax=1,
ylabel style={font=\color{white!15!black},font=\footnotesize},
ylabel={{Pr(position error $<\epsilon$)}},
axis background/.style={fill=white},
xmajorgrids,
xminorgrids,
ymajorgrids,
legend style={at={(1,0)}, anchor=south east, ,font=\footnotesize, legend cell align=left, align=left, draw=white!15!black, legend columns=1}
]

\addplot [color=blue, mark=o, line width=0.7pt, mark options={solid, blue}]
  table[row sep=crcr]{%
0.01	0.00624739691795084\\
0.0114975699539774	0.0187421907538525\\
0.0132194114846603	0.0336318200749688\\
0.0151991108295293	0.0622657226155768\\
0.0174752840000768	0.119533527696793\\
0.0200923300256505	0.215535193669304\\
0.0231012970008316	0.345689296126614\\
0.0265608778294669	0.495002082465639\\
0.0305385550883342	0.604019158683882\\
0.0351119173421513	0.68981674302374\\
0.0403701725859655	0.756351520199917\\
0.0464158883361278	0.802582257392753\\
0.0533669923120631	0.838400666389005\\
0.0613590727341317	0.870158267388588\\
0.0705480231071865	0.895564348188255\\
0.0811130830789687	0.912536443148688\\
0.093260334688322	0.928571428571429\\
0.107226722201032	0.94283631820075\\
0.123284673944207	0.952519783423573\\
0.141747416292681	0.961057892544773\\
0.162975083462064	0.967930029154519\\
0.187381742286038	0.97240733027905\\
0.215443469003188	0.97542690545606\\
0.247707635599171	0.978550603915035\\
0.28480358684358	0.981570179092045\\
0.327454916287773	0.985735110370679\\
0.376493580679247	0.988234069137859\\
0.432876128108306	0.988546438983757\\
0.497702356433211	0.988858808829654\\
0.572236765935022	0.98896293211162\\
0.657933224657568	0.98896293211162\\
0.756463327554629	0.98896293211162\\
0.869749002617783	0.98896293211162\\
1	0.98896293211162\\
};
\addlegendentry{4-RIS}

\addplot [color=green, mark=diamond, line width=0.7pt, mark options={solid, green}]
  table[row sep=crcr]{%
0.01	0.00239483548521446\\
0.0114975699539774	0.00770512286547276\\
0.0132194114846603	0.0150978758850479\\
0.0151991108295293	0.026655560183257\\
0.0174752840000768	0.0485214493960849\\
0.0200923300256505	0.0842357351103706\\
0.0231012970008316	0.132444814660558\\
0.0265608778294669	0.193773427738442\\
0.0305385550883342	0.259995835068721\\
0.0351119173421513	0.335381091211995\\
0.0403701725859655	0.410662224073303\\
0.0464158883361278	0.484485630987089\\
0.0533669923120631	0.552478134110787\\
0.0613590727341317	0.612244897959184\\
0.0705480231071865	0.666701374427322\\
0.0811130830789687	0.711786755518534\\
0.093260334688322	0.750624739691795\\
0.107226722201032	0.789775093710954\\
0.123284673944207	0.81820074968763\\
0.141747416292681	0.843606830487297\\
0.162975083462064	0.865056226572262\\
0.187381742286038	0.882444814660558\\
0.215443469003188	0.900249895876718\\
0.247707635599171	0.915556018325698\\
0.28480358684358	0.92877967513536\\
0.327454916287773	0.938775510204082\\
0.376493580679247	0.947938359017076\\
0.432876128108306	0.95637234485631\\
0.497702356433211	0.962099125364432\\
0.572236765935022	0.966784673052895\\
0.657933224657568	0.970012494793836\\
0.756463327554629	0.972199083715119\\
0.869749002617783	0.975010412328197\\
1	0.978550603915035\\
};
\addlegendentry{3-RIS}

\addplot [color=black, dashed, line width=0.7pt, mark=+, mark options={solid, black}]
  table[row sep=crcr]{%
0.01	0.0185339441899208\\
0.0114975699539774	0.0302998750520617\\
0.0132194114846603	0.0486255726780508\\
0.0151991108295293	0.0845481049562682\\
0.0174752840000768	0.128800499791753\\
0.0200923300256505	0.180862140774677\\
0.0231012970008316	0.228654727197001\\
0.0265608778294669	0.269783423573511\\
0.0305385550883342	0.310807996668055\\
0.0351119173421513	0.35079133694294\\
0.0403701725859655	0.393898375676801\\
0.0464158883361278	0.439296126613911\\
0.0533669923120631	0.485110370678884\\
0.0613590727341317	0.522594752186589\\
0.0705480231071865	0.559975010412328\\
0.0811130830789687	0.593815077051229\\
0.093260334688322	0.628904623073719\\
0.107226722201032	0.661703456892961\\
0.123284673944207	0.693148688046647\\
0.141747416292681	0.722094960433153\\
0.162975083462064	0.749895876718034\\
0.187381742286038	0.778113286130779\\
0.215443469003188	0.806538942107455\\
0.247707635599171	0.830279050395669\\
0.28480358684358	0.846730528946272\\
0.327454916287773	0.868075801749271\\
0.376493580679247	0.887338608912953\\
0.432876128108306	0.903269471053728\\
0.497702356433211	0.918159100374844\\
0.572236765935022	0.930758017492711\\
0.657933224657568	0.940649729279467\\
0.756463327554629	0.950333194502291\\
0.869749002617783	0.959079550187422\\
1	0.96563931695127\\
};
\addlegendentry{2-RIS (L), 2D}

\addplot [color=black, dashed, line width=0.7pt, mark=triangle, mark options={solid, black}]
  table[row sep=crcr]{%
0.01	0.00218658892128276\\
0.0114975699539774	0.00489379425239489\\
0.0132194114846603	0.00853810912119946\\
0.0151991108295293	0.0156184922948771\\
0.0174752840000768	0.0255102040816326\\
0.0200923300256505	0.0417534360683048\\
0.0231012970008316	0.0598708871303624\\
0.0265608778294669	0.085381091211995\\
0.0305385550883342	0.114743856726364\\
0.0351119173421513	0.150562265722616\\
0.0403701725859655	0.192940441482715\\
0.0464158883361278	0.237921699291962\\
0.0533669923120631	0.29487713452728\\
0.0613590727341317	0.347563515201999\\
0.0705480231071865	0.40295710120783\\
0.0811130830789687	0.454498125780925\\
0.093260334688322	0.506663890045814\\
0.107226722201032	0.555393586005831\\
0.123284673944207	0.598188254893794\\
0.141747416292681	0.637442732194919\\
0.162975083462064	0.676384839650146\\
0.187381742286038	0.713348604748022\\
0.215443469003188	0.745835068721366\\
0.247707635599171	0.770720533111204\\
0.28480358684358	0.795085381091212\\
0.327454916287773	0.819241982507289\\
0.376493580679247	0.845585172844648\\
0.432876128108306	0.865993336109954\\
0.497702356433211	0.88765097875885\\
0.572236765935022	0.905143690129113\\
0.657933224657568	0.918159100374844\\
0.756463327554629	0.928675551853394\\
0.869749002617783	0.937734277384423\\
1	0.944294044148272\\
};
\addlegendentry{2-RIS (L), Sync}

\addplot [color=red, mark=square, line width=0.7pt, mark options={solid, red}]
  table[row sep=crcr]{%
0.01	0.000728862973760958\\
0.0114975699539774	0.00249895876718031\\
0.0132194114846603	0.00416493127863393\\
0.0151991108295293	0.00801749271137031\\
0.0174752840000768	0.0144731361932529\\
0.0200923300256505	0.0225947521865889\\
0.0231012970008316	0.0359225322782174\\
0.0265608778294669	0.0482090795501874\\
0.0305385550883342	0.0630987088713036\\
0.0351119173421513	0.078508954602249\\
0.0403701725859655	0.10006247396918\\
0.0464158883361278	0.123177842565598\\
0.0533669923120631	0.145668471470221\\
0.0613590727341317	0.176176593086214\\
0.0705480231071865	0.219804248229904\\
0.0811130830789687	0.27582257392753\\
0.093260334688322	0.331840899625156\\
0.107226722201032	0.386609745939192\\
0.123284673944207	0.43606830487297\\
0.141747416292681	0.485110370678884\\
0.162975083462064	0.532069970845481\\
0.187381742286038	0.573302790503957\\
0.215443469003188	0.608808829654311\\
0.247707635599171	0.644002498958767\\
0.28480358684358	0.679820907955019\\
0.327454916287773	0.713660974593919\\
0.376493580679247	0.744064972927947\\
0.432876128108306	0.773531861724282\\
0.497702356433211	0.800812161599334\\
0.572236765935022	0.820491461890879\\
0.657933224657568	0.840170762182424\\
0.756463327554629	0.85683048729696\\
0.869749002617783	0.870991253644315\\
1	0.883694294044148\\
};
\addlegendentry{2-RIS (P)}

\addplot [color=black, line width=0.7pt]
  table[row sep=crcr]{%
0.01	0.000832986255726809\\
0.0114975699539774	0.00197834235735106\\
0.0132194114846603	0.00333194502290712\\
0.0151991108295293	0.00572678050812159\\
0.0174752840000768	0.0108288213244482\\
0.0200923300256505	0.0166597251145356\\
0.0231012970008316	0.0233236151603499\\
0.0265608778294669	0.0318617242815493\\
0.0305385550883342	0.0412328196584756\\
0.0351119173421513	0.0518533944189921\\
0.0403701725859655	0.0652852977925864\\
0.0464158883361278	0.0841316118284048\\
0.0533669923120631	0.106101624323199\\
0.0613590727341317	0.132132444814661\\
0.0705480231071865	0.160453977509371\\
0.0811130830789687	0.187213660974594\\
0.093260334688322	0.217305289462724\\
0.107226722201032	0.248229904206581\\
0.123284673944207	0.282382340691379\\
0.141747416292681	0.316743023740108\\
0.162975083462064	0.353082049146189\\
0.187381742286038	0.392232403165348\\
0.215443469003188	0.428467305289463\\
0.247707635599171	0.467825905872553\\
0.28480358684358	0.505726780508122\\
0.327454916287773	0.542586422324032\\
0.376493580679247	0.576322365680966\\
0.432876128108306	0.608600583090379\\
0.497702356433211	0.641399416909621\\
0.572236765935022	0.669304456476468\\
0.657933224657568	0.699083715118701\\
0.756463327554629	0.725947521865889\\
0.869749002617783	0.752082465639317\\
1	0.77207413577676\\
};
\addlegendentry{2-RIS (L)}

\end{axis}
\end{tikzpicture}%
\vspace{-1.1 cm}
\caption{CDF of the $\text{PEB}_\text{R}$ for both UEs at different locations inside a $\unit[7\times 7]{m^2}$ area. RIS (L) indicates RISs are located in an L-shape, as shown in Fig.~\ref{fig_multiple_RISs}~(a), and RIS (P) indicates RISs are located in parallel, as shown in Fig.~\ref{fig_multiple_RISs}~(b).}
\label{fig_multiple_RIS_cdf}
\vspace{-0.cm}
\end{figure}
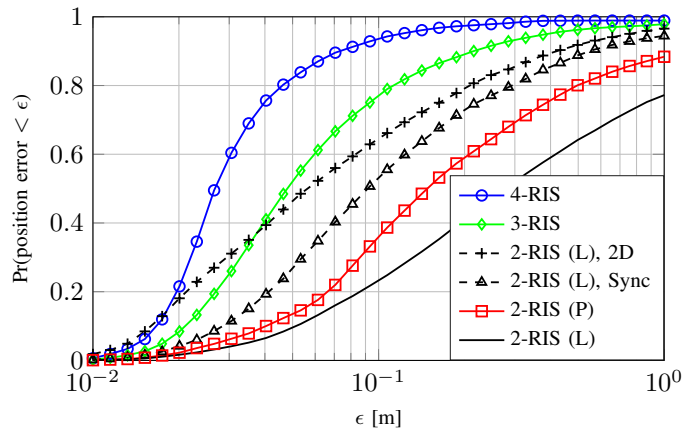

Considering {the localization performance highly depends on the location of RIS anchors, we can not have a general analytical solution for different numbers of RISs.} Here, we {numerically} evaluate the effect of the number of RISs on positioning with four candidate anchor positions ($[-4, 0, 2]^\top, [0, 4, 2]^\top, [4, 0, 2]^\top, [0, -4, 2]^\top$) covering an $\unit[8\times 8]{m^2}$ area. The RIS tilt angles are set as $30^\circ$ pointing down, and we can see that RIS orientation affects the blind area (see Fig.~\ref{fig_multiple_RISs} (a) vs. Fig.~\ref{fig_heatmap} (c)-1, and Fig.~\ref{fig_multiple_RISs} (b) vs. Fig.~\ref{fig_heatmap} (a)-1). In general, more RISs can increase positioning coverage; however, {the improvement depends on the position of UEs and RISs}. {In addition, }if the same orthogonal strategy is implemented, more blocks are needed, increasing difficulties in coordinating between these RISs and channel parameter estimation. To support this statement in a more general scenario, we assume that the TX and RX UEs can be located at grids with $x\in \{-3, -2, \ldots, 3\}\ \text{m}$, $y\in \{-3, -2, \ldots, 3\}\ \text{m}$, and $z\in \{0, 0.5\}\ \text{m}$. We further assume both UEs cannot be located at the same place, and hence a total number of ${{7\times7\times 2}\choose{2}} = 4753$ TX-RX pairs can be evaluated. The \ac{cdf} of the $\text{PEB}_\text{R}$ for all the scenarios in Fig.~\ref{fig_multiple_RISs} is shown in Fig.~\ref{fig_multiple_RIS_cdf}. The CDFs for the RIS (L) scenario with known height (black dashed curve with cross markers) and known clock offset (black dashed curve with triangle markers) are also plotted, which validates the suggested solutions in reducing the blind areas.

\subsubsection{Summary}
We have shown that blind areas exist in the problem of multi-RIS-enabled 3D sidelink positioning with a limited number of RISs (i.e., two), and we have provided several ways to mitigate the effect, namely, round-trip positioning to remove clock bias, consider geometric constraints, and adopt more RISs. These discussions also open new directions for offline and online system optimization. The offline deployment of the anchors needs to consider the TX and RX position probability (e.g., the vehicles can only drive on the road with a certain movement model), as well as the surrounding environment map (e.g., where RISs can be installed). The online optimization needs to take advantage of the prior information and consider when, and to which UE, to trigger a positioning process, as the positioning performance may not meet the positioning performance requirements in the blind areas.

\section{Conclusion}
\label{sec_conclusion}
In this work, we have formulated and solved the multi-RIS-enabled 3D sidelink positioning problem. 
In this problem, with the assistance of at least two RISs, the absolute positions of two unsynchronized UEs can be estimated via a one-way sidelink communication in the absence of BSs. Channel parameter estimation and positioning algorithms are developed and benchmarked by the derived CRBs. We discussed the effect of multipath on positioning performance and found the impact on the RIS channels is more significant. We also evaluated the benefit of RIS profile design with prior information to boost positioning performance. Most importantly, we have shown that blind areas exist in RIS-enabled sidelink positioning problems with interpretations. Several solutions can be considered to reduce the effect of blind areas, such as utilizing round-trip communication to remove clock offset, adding geometric constraints to reduce the number of unknowns, and adopting more RISs to increase positioning coverage. However, this work is just the starting point for sidelink positioning with simplified scenarios and channel models. Further directions can consider the high-mobility scenario with the Doppler effect and more accurate channel models that account for more features, such as the near-field effect and beam squint effect. 
{Moreover, machine learning algorithms can be developed to reduce the algorithm complexity of RIS profile design and weighting coefficients optimization in the positioning algorithm when more RISs are involved.}

\begin{appendices}

\section{{Equivalent Convex Reformulation of \eqref{eq_opt_formulation_2}}}\label{app_conv_form} 

{To obtain a convex optimization problem from \eqref{eq_opt_formulation_2}, we first express the PEBs in \eqref{eq:PEBR} explicitly as a function of $\deltav$: $\mathrm{PEB^2_\text{R}}(\pv_{\text{T}, \tilde g}, \pv_{\text{R}, \tilde g}, \Xim_1, \ldots,  \Xim_L|\deltav) = \trace([\bm{\mathcal{I}}({\sv_{\tilde g}})^{-1}]_{4:6, 4:6})$,
where, using \eqref{eq:FIM_measurement} and \eqref{eq_crb_sidelink}, 
\begin{align} 
\label{eq_Isvg_}
    \bm{\mathcal{I}}(\sv_{\tilde g}) &= \Jm_\mathrm{S} \bm{\mathcal{I}}({\etav_{\tilde g}}) \Jm_\mathrm{S}^\top ~, \\
    \bm{\mathcal{I}}({\boldsymbol\eta_{\tilde g}}) 
    &= \frac{2}{\sigma_n^2}\sum^{G}_{g=1} \sum^K_{k=1}\mathrm{Re}\left\{
    \left(\frac{\partial{\mu}_{k,g}(\tilde g)}{\partial{\boldsymbol\eta}_{\tilde g}}\right)^{\mathsf{H}} 
    \left(\frac{\partial{\mu}_{k,g}(\tilde g)}{\partial{\boldsymbol\eta}_{\tilde g}}\right)\right\} ~,
    \label{eq:FIM_measurement_r1_} \\ \label{eq:FIM_measurement_nu_r1_}
    \mu_{k,g}(\tilde g) &= [\Ym_{\text{U}}(\tilde g)]_{k,g} + [\Ym_{\text{R}}(\tilde g)]_{k,g} ~.
\end{align}
Plugging (6)--(8) into \eqref{eq:FIM_measurement_nu_r1_}, we obtain $\mu_{k,g}(\tilde g) = [\Tm(\tilde g)]_{k,g} \sqrt{P} x_k \delta_g$,
where $\Tm(\tilde g) \triangleq \Hm_{\text{U}} + \Hm_{\text{U,MP}} + \sum_{\ell=1}^{L}(\Hm_{\text{R},\ell} + \Hm_{\text{R,MP},\ell})$.}
%
{We now insert this into \eqref{eq:FIM_measurement_r1_} to obtain $\bm{\mathcal{I}}({\boldsymbol\eta_{\tilde g}}) 
    = \sum^{G}_{g=1} \gamma_g \Om_g(\tilde g) $
for some known matrix $\Om_g(\tilde g) $,  where $\gamma_g \triangleq \delta_g^2$. 
Substituting this into \eqref{eq_Isvg_} yields
\begin{align}
    \bm{\mathcal{I}}({\sv_{\tilde g}}) 
    &= \sum^{G}_{g=1} \gamma_g \Sm_g(\tilde g) ~,
    \label{eq:FIM_measurement_r1_v3_} 
\end{align}
where $\Sm_g(\tilde g) \triangleq \Jm_\mathrm{S} \Om_g(\tilde g) \Jm_\mathrm{S}^\top $. Performing change of variables $\bm\gamma = [\gamma_1 \, \ldots \, \gamma_G]^\top$, the problem \eqref{eq_opt_formulation_2} can be expressed as
\begin{align}
  \min_{\bm\gamma \in \mathbb{R}^G} 
  \frac{3}{\tilde G}\sum_{\tilde g = 1}^{\tilde G/3} & \trace([\bm{\mathcal{I}}({\sv_{\tilde g}})^{-1}]_{4:6, 4:6}) 
  \label{eq_opt_formulation_3_r1_}
  ~ & \mathrm{s.t.} ~~  \boldone^\top \bm\gamma = G ~.
\end{align}
Introducing auxiliary variables $u_{\tilde g}$, \eqref{eq_opt_formulation_3_r1_} can be recast as
\begin{subequations}\label{eq_problem_all_sub_}
\begin{align}
  \min_{\bm\gamma \in \mathbb{R}^G, \{u_{\tilde g}\}} 
  &~~ \frac{3}{\tilde G}\sum_{\tilde g = 1}^{\tilde G/3} u_{\tilde g}
  \label{eq_opt_formulation_4_r1_}
  \\ \label{eq_const_g_}
  ~~  \mathrm{s.t.} &~~  \boldone^\top \bm\gamma = G  \,, \,
   \trace([\bm{\mathcal{I}}({\sv_{\tilde g}})^{-1}]_{4:6, 4:6}) \leq u_{\tilde g}, \forall \tilde g ~.
\end{align}
\end{subequations}
From \cite[Ch.~7.5.2]{boyd2004convex}, each constraint in \eqref{eq_const_g_} can be equivalently reformulated as a linear matrix inequality (LMI) constraint, leading to (using \eqref{eq:FIM_measurement_r1_v3_})
\begin{subequations}\label{eq_problem_all_sub2_}
\begin{align}
  \min_{\bm\gamma \in \mathbb{R}^G, \{u_{\tilde g}\}, \{u_{\tilde g,\ell} \}} 
  &~~ \frac{3}{\tilde G}\sum_{\tilde g = 1}^{\tilde G/3} u_{\tilde g}
  \label{eq_opt_formulation_5_r1_}
  \\
  ~~  \mathrm{s.t.} &~~  \boldone^\top \bm\gamma = G \notag ~,
  \\ \nonumber
  & ~~ \begin{bmatrix}
       \sum^{G}_{g=1} \gamma_g \Sm_g(\tilde g) & \ev_\ell \\
       \ev_\ell^\top & u_{\tilde g,\ell}
   \end{bmatrix} \succeq 0, \ell = 4,5,6,
   \\ \label{eq_const_g2_} &~ \sum_{\ell = 4}^{6} u_{\tilde g,\ell} \leq  u_{\tilde g}, \forall \tilde g ~,
\end{align}
\end{subequations}
where $u_{\tilde g,\ell}$ denote the newly introduced auxiliary variables. The problem \eqref{eq_problem_all_sub2_}, which is equivalent to \eqref{eq_opt_formulation_2}, is a convex semidefinite programming (SDP) problem \cite[Ch.~4.6.2]{boyd2004convex}.} 

\end{appendices}


\bibliographystyle{IEEEtran}
\bibliography{IEEEabrv, ref}

\end{document}